\documentclass[prc,twocolumn,amsmath,amssymb,superscriptaddress,nofootinbib,preprintnumbers]{revtex4-1}
\usepackage{url,color,colordvi,epsfig,pstricks}
\usepackage[colorlinks=true, citecolor=blue, filecolor=blue, linkcolor=blue, urlcolor=blue, pdftex]{hyperref}
\usepackage{graphicx,bm,dcolumn}
\usepackage{hyperref,slashed,lineno}
\usepackage{physics}
\usepackage{inputenc}
\usepackage{soul}

\hypersetup{
    colorlinks=true, 
    linktoc=all, 
    linkcolor=blue, 
    citecolor=blue}

\def\Fermilab{Theory Division, Fermilab, P.O. Box 500, Batavia, IL 60510, USA}
\def\IFIC{Instituto de Física Corpuscular (IFIC), CSIC‐Universitat de València, Spain}

\begin{document}
\preprint{NT@UW-26-10}

\title{Computing neutrino cross sections from Euclidean responses}
\author{A.~Nikolakopoulos}\email{anikolak@uw.edu}
\affiliation{Department of Physics, University of Washington, Seattle WA, USA}
\author{N.~Rocco}\email{rocco@ific.us.es}
\affiliation{\IFIC}
\affiliation{\Fermilab}

\begin{abstract}
Energy integrated neutrino cross sections are integrals of nuclear responses weighted with kinematic prefactors.
We decompose the prefactors into a limited set of functions of energy transfer and show the relevant integrals are the moments of the responses, and integrals weighted with $1/(a+\omega)^n$ with $n\leq 2$. 
These can be directly obtained from the Euclidean response, avoiding the need for inversion of the Laplace transform.
As a proof of concept we study the procedure with toy-model responses for the quasielastic peak. 
We show that the different contributions can be straightforwardly organized in terms of relative importance, and how flux-averaged cross sections can be obtained.
Using a realistic model for the response and numerical uncertainty we show that it is feasible to obtain the required integrals from the Euclidean response, with large uncertainties only for the third moment. 
Due to kinematic restrictions, the integrals contain contributions from the unphysical region for neutrino scattering, coming from high-momentum nucleons.
We show that (in the absence of two-body currents) robust corrections for this contamination are obtained from the single-nucleon momentum distribution.
These results present an opportunity to compute certain neutrino cross sections with ab-initio methods with controlled uncertainties.
\end{abstract}

\maketitle

\section{Introduction}
As neutrino physics moves into a precision era, the need for reliable neutrino–nucleus cross sections becomes more acute~\cite{NUSTECWP, Ruso:2022qes}.
On the other hand, current and future neutrino experiments are sources of unique data that can elucidate electroweak properties of nucleons and nuclei.
In recent years, ab initio many-body methods have delivered accurate predictions for a wide class of nuclear electroweak observables which may be confronted with data, including electromagnetic dipole responses, muon capture rates, and neutrino–nucleus cross sections relevant to oscillation experiments~\cite{RevModPhys.87.1067, LEIDEMANN2013158, Lovato16,PhysRevX.10.031068,Nikolakopoulos:2023zse, Sobczyk:2021dwm,Sobczyk:2023sxh,Acharya:2024xah,Miorelli:2016qbk,Birkhan:2016qkr}.
Here ab initio refers to systematically improvable many-body calculations that start from nucleon–nucleon (and multi-nucleon) interactions and consistent electroweak current operators.

In inclusive electroweak interactions, nuclear response functions are the relevant ingredients.
In many approaches, rather than resolving the full energy dependence of the responses, one computes their convolution with a known kernel. 
In Green’s Function Monte Carlo calculations, this is typically done using the Laplace kernel, leading to the Euclidean response (an imaginary-time correlator)~\cite{RevModPhys.87.1067}.
Early GFMC studies used the Euclidian response to quantify electroweak strength and the influence of two-body-currents~\cite{Lovato14,PhysRevC.65.024002, PhysRevC.49.R2880}.
In coupled-cluster~\cite{Hagen:2013nca} and hyperspherical-harmonics approaches~\cite{Barnea:1999be}, a common alternative is the Lorentzian kernel, giving the Lorentz Integral Transform (LIT)~\cite{EFROS1994130, LEIDEMANN2013158}.
Recovering the physical response requires an inversion of the corresponding integral transform, which is intrinsically ill-posed. 
That is, once realistic (numerical) uncertainties are included small fluctuations in the transform can be amplified into large variations in the reconstructed responses unless strong regularization or prior assumptions are imposed.

Maximum-entropy (MaxEnt) methods have long been the workhorse for extracting energy-transfer response functions from imaginary-time (Euclidean) correlators in ab initio calculations~\cite{Lovato:2017cux,Lovato16}. MaxEnt analytic continuation was employed to reconstruct the energy-dependence of inclusive responses, e.g. in the GFMC calculation of Refs.~\cite{Lovato:2017cux, PhysRevC.91.062501, Lovato16, PhysRevX.10.031068}.
A well-known limitation of MaxEnt is that propagating the statistical uncertainties of the Euclidean data into reliable, point-wise uncertainty bands for the inverted response remains challenging and often method-dependent.
More recently, methods leveraging machine-learning techniques have been proposed. A physics-informed neural-network architecture was introduced in Ref.~\cite{raghavan_machine_2021}, and improved reconstruction relative to MaxEnt was demonstrated in regimes relevant for neutrino physics. 
Follow-up work focused explicitly on uncertainty quantification~\cite{raghavan_uncertainty-quantification-enabled_2024}.
For the LIT, a Chebyshev-polynomial reconstruction of spectral densities was developed in~\cite{Sobczyk:2021ejs}, providing an alternative route to stabilize the inversion. Very recently, an enhanced MaxEnt procedure tailored to LIT inversions was introduced alongside a neural-network–quantum-state + LIT framework~\cite{Parnes:2025seu}, illustrating how methodological refinements can be coupled to new many-body representations. 
Finally, lattice-QCD studies have advanced rigorous analytic-continuation tools, including Nevanlinna–Pick interpolation and moment-problem techniques, with recent overviews and applications appearing in Refs.~\cite{Jay:2025dzl,Salg:2025now}.
These developments underscore that inversion problems are currently a major focus in the field.

In this work, we provide a complementary perspective. 
We show that energy-integrated neutrino cross sections at fixed momentum transfer can be expressed in terms of a small set of energy-weighted integrals of the nuclear responses. Crucially, these integrated quantities may be extracted directly from Euclidean responses, thereby avoiding a full inversion of the Laplace transform.
In this way, the propagation of statistical and theoretical uncertainties from the Euclidean data to the cross sections is straightforward. 
Moreover, under the assumption that the response is dominated by a quasielastic peak, the relevance of different integrated contributions to the cross section can be naturally organized.

We show that the cross sections are linear combinations of integrals of the responses weighted with polynomials (i.e. the moments of the responses) and functions $(a+x)^{-n}$. We show how these can be obtained from Euclidean response data, under assumptions that we discuss in detail in the main text.
We validate the method first in a controlled setting using a Gaussian toy model.
We then then assess its feasibility in a more realistic framework. To do this, we use Euclidean responses computed in the plane-wave impulse approximation with a realistic nuclear spectral function, and include numerical uncertainties representative of GFMC calculations.

In section~\ref{sec:method} the basic idea is presented and we derive expressions for cross sections integrated over outgoing lepton energy and for flux-averaged cross sections in terms of weighed integrals of the Euclidean responses. In section~\ref{sec:toymodel} we use a toy model for the nuclear responses as a proof of concept, and study the relative importance of different contributions.
In section~\ref{sec:SF_integrals} we use a realistic model for the Euclidian response and its numerical uncertainty. We show how a correction for contamination to the integrals from the response in the unphysical region might be obtained from the single-nucleon momentum distribution.
In section~\ref{sec:considerations} we discuss complications due to two-body currents, and nucleon form factors.
Conclusions and a summary are given in section~\ref{sec:conclusion}.
\section{Method}
\label{sec:method}
In four-point Fermi theory the inclusive cross section for neutrino nucleus scattering can be expressed in terms of five nuclear response functions~\cite{walecka04}
\begin{align}
\label{eq:CS_wq_vR}
\frac{\mathrm{d}\sigma(E)}{\mathrm{d}\omega \mathrm{d} q} &= \frac{G_F^2 E_f q}{E} \left[ v_{CC} R_{CC} + v_{CL}R_{CL} + v_{LL} R_{LL} \right. \\ \nonumber 
&\left. + v_T R_T + v_{T^\prime} R_{T^\prime} \right].
\end{align}
Here $E, E_f$ are the energies of the initial and final-state lepton, and $\omega = E - E_f$, $q = \lvert \mathbf{k} - \mathbf{k}_f \rvert$ the energy and momentum transferred to the nucleus respectively.
The lepton factors $v_i(E,E_f,q)$ depend on the incoming energy, while the nuclear responses $R(\omega,q)$ only depend on $\omega$ and $q$.
The lepton factors can be computed straightforwardly at tree level, while the nuclear responses are computed by many-body methods.
In Greens Function Monte Carlo (GFMC), these responses are not computed directly.
Instead the GFMC provides the Euclidean response, i.e. the Laplace transforms of the nuclear responses
\begin{equation}
E_i(\tau,q) \equiv \mathcal{L}\left[R_i(\omega,q) \right](\tau,q) = \int_0^{\infty} R_i(\omega, q) e^{-\tau\omega} \mathrm{d}\omega.
\end{equation}
To recover the responses needed in Eq.~(\ref{eq:CS_wq_vR}), the Laplace transform has to be inverted.

We point out that there are quantities of interest which may be computed directly from the Euclidean response, avoiding the inversion procedure.
These are energy-averaged cross sections which can be written as an integral over $\omega$ of the responses.
First, we consider the flux-averaged cross section
\begin{align}
\label{eq:CS_FF_Eq}
&\left\langle \frac{\mathrm{d}\sigma}{\mathrm{d}E_f \mathrm{d}q} \right\rangle = G_F^2 E_f q \\ \nonumber
&\times \int \mathrm{d}\omega~\frac{\phi(\omega + E_f)}{E_f+\omega}\sum_i v_i(\omega+E_f,E_f,q) R_i(\omega,q),
\end{align}
where $\phi(E)$ is the incoming neutrino flux. 
Second, the cross section integrated over the final lepton energy
\begin{equation}
\label{eq:CS_diff_q}
    \frac{\mathrm{d}\sigma(E)}{\mathrm{d}q} = \frac{G_F^2 q}{E}\int \mathrm{d}\omega (E - \omega) \sum_i v_i(E,E-\omega,q) R(\omega,q)\, .
\end{equation}
These cross sections are given by weighted integrals over $\omega$ of the nuclear response functions.
The known kinematic prefactors in Eqs.~(\ref{eq:CS_FF_Eq} - \ref{eq:CS_diff_q}) may be decomposed into a minimal set of functions of $\omega$.
The cross sections are then only sensitive to the integrals of the response weighted with these functions; the exact $\omega$-dependence of the responses is not required.
Some of these integrals might be computed directly from the Euclidean response.
Indeed, the Laplace transform satisfies
\begin{equation}
\label{eq:laplace_fold}
\int_0^{\infty} f(\omega) R(\omega) \mathrm{d}\omega = \int_{0}^{\infty} \mathcal{L}\left[R \right](\tau) \mathcal{L}^{-1} \left[f \right](\tau) \mathrm{d}\tau
\end{equation}
where $\mathcal{L}^{-1}$ is the inverse Laplace transform\footnote{
This follows simply by inserting $f(\omega) = \mathcal{L}\left[\mathcal{L}^{-1} \left[f(\omega) \right]\right]$ if the order of integration can be exchanged.}.
Hence, using the inverse Laplace transform of the known kinematic prefactors, Eqs.~(\ref{eq:CS_FF_Eq})~and~(\ref{eq:CS_diff_q}) can be computed as integrals over the Euclidean response if the upper bounds of the integrals can be taken to infinity\footnote{This assumption is studied in Sec.~\ref{sec:SF_integrals}}.
In the following we give the decomposition of the kinematic prefactors in terms of simple functions of $\omega$ for charged-current scattering for massless leptons\footnote{Lepton mass corrections are discussed in Sec.~\ref{sec:leptonmass} and pose no complications.}.

\subsection{Unit flux}
\label{sec:unitflux_bounds}
We consider the energy-averaged cross section, that is Eq.~(\ref{eq:CS_FF_Eq}) for a uniform flux.
These results provide the building block for more complicated cases with additional energy-dependent prefactors, e.g. the flux-averaged cross section.
The energy averaged cross section for fixed $E_f$ and $q$ is
\begin{align}
&\left\langle \frac{\mathrm{d}\sigma}{\mathrm{d}E_f\mathrm{d}q} \right\rangle = G^2 q \nonumber \\
&\times\int_0^{q}\mathrm{d}\omega  \frac{E_f}{E_f+\omega} \sum_{i} v_i(\omega+E_f, E_f, q) R_i(\omega,q).
\label{eq:CS_unitflux}
\end{align}
The functions $v_i$ are usually written in terms of $E,~E_f,$ and the lepton scattering angle $\cos\theta_f$.
They are given in Eqs.~(\ref{eq:VCC}-\ref{eq:VTp}) for the case of massless final-state leptons. This case is considered in this work, but as shown in Sec.~\ref{sec:leptonmass} lepton mass terms pose no issues.
At fixed $q$ we then have
\begin{equation}
\label{eq:costheta_wEf}
\cos\theta(q,\omega,E_f) = \frac{q^2 - (\omega + E_f)^2 - E_f^2}{-2(\omega+E_f)E_f}.
\end{equation}
The requirement of $\cos\theta > -1$ leads to $E_f^{\mathrm{min}} = q/2$. This is the minimum lepton energy accessible at fixed $q$ (for $\omega=0$). The maximum value of the accessible lepton energy is $E^{max}_f=q$.
The limit $\cos\theta < 1$ implies that $\omega^{\mathrm{max}} = q$, which gives the upper bound in the integral of Eq.~(\ref{eq:CS_unitflux}).

Each of the kinematic prefactors $\frac{E_f}{E_f + \omega} v_i$ can be written as a linear combination of at most 5 different functions of $\omega$. These are $\omega^n$ for $n \in \{0,1,2\}$ and 
\begin{equation}
    \mathcal{E}^n(\omega) \equiv \frac{1}{(\omega+E_f)^n}
\end{equation}
for $n \in \{1,2\}$. The explicit expressions are given in Eqs.~(\ref{eq:VCC_DE}-\ref{eq:VTp_DE}).
Their inverse Laplace transforms are $\mathcal{L}^{-1}\left[ \omega^n \right](\tau) = \delta^{(n)}(\tau)$, where the superscript $^{(n)}$ denotes the $n$-th derivative, and $\mathcal{L}^{-1}\left[ \mathcal{E}^n\right](\tau) = \tau^{n-1}e^{-E_f\tau}$.
Thus the energy-averaged cross section can be written in terms of at most five integrals of each of the responses.
If we, \emph{for now}, assume that the upper limit of the integral in Eq.~(\ref{eq:CS_unitflux}) can be taken arbitrarily large these are the moments of the response
\begin{equation}
\label{eq:Int_Moments}
I_i\left[ \omega^n \right] \equiv \int_0^{\infty} \mathrm{d}\omega~\omega^n R_i(\omega,q) = (-1)^n E_i^{(n)}(0),
\end{equation}
given by the derivatives of the Euclidean response (the moment-generating function) at $\tau = 0$.
And the energy-dependent integrals
\begin{align}
\label{eq:intE}
I_i\left[\mathcal{E}^1 \right] &\equiv \int_0^{\infty} \mathrm{d}\omega~\frac{R_i(\omega,q)}{E_f+\omega} =  \int_0^{\infty} \mathrm{d}\tau E_i(\tau) e^{-E_f\tau}, \\
I_i\left[\mathcal{E}^2 \right] &\equiv \int_0^{\infty} \mathrm{d}\omega~\frac{R_i(\omega,q)}{(E_f+\omega)^2} = \int_0^{\infty} \mathrm{d}\tau E_i(\tau) \tau e^{-E_f\tau}. \label{eq:intE2}
\end{align}
The energy-integrated cross section is a linear combination of these integrals,
\begin{equation}
    \label{eq:defVE_D}
    \left\langle\frac{\mathrm{d}\sigma}{\mathrm{d}E_l\mathrm{d}q}\right\rangle = G_F^2 q \left( V^{\omega}_{ij} I_i\left[\omega^j\right] + V_{ij}^{E} I_i\left[\mathcal{E}^j\right] \right),
\end{equation}
where summation over $i$ and $j$ is implied, and $V^{\omega}$ and $V^{E}$ are matrices containing to the expansion coefficients for the $i$-th lepton factor in terms of the functions $\omega^j$ and $\mathcal{E}^j$ respectively. These matrices can be directly read off from Eqs.~(\ref{eq:VCC_DE}-\ref{eq:VTp_DE}).

\subsection{Flux-averaging}
\label{sec:fluxfolding}
For this case one needs the decomposition of the functions
\begin{equation}
    \frac{E_f}{E_f+\omega} v_i(E_f+\omega, E_f, q) \phi(E_f+\omega),
\end{equation}
where $\phi(E)$ is the incoming neutrino flux.
The flux is generally not known analytically. It is sufficient to approximate the flux by an analytic function $f(E)$ in the energy region $E \in \left[E_f, E_f+q \right]$, which is the physical region probed in Eq.~(\ref{eq:CS_unitflux}). The accuracy of the approximation will depend on the shape of the flux in this energy window and will generally improve as $q$ becomes smaller, since a  narrower window needs to be considered.
The choice of functions is arbitrary; the only requirement is that $\mathcal{L}^{-1}\left[ f(\omega+E_f) \omega^n \right](\tau)$ and $\mathcal{L}^{-1}\left[ f(\omega+E_f) \mathcal{E}^n \right](\tau)$ are well behaved (e.g. the $f(\omega)$ have no poles in the right-hand $\omega$-plane) so that the resulting integrals of the Euclidean response can be computed.

In the following, we will consider a second-order polynomial to approximate the flux for simplicity.
In this case it is particularly straightforward to express the flux-averaged cross section in terms of the same integrals $I_i\left[\omega^j\right]$ and $I_i\left[ \mathcal{E}^j\right]$.
Since we have $\omega \mathcal{E}^n(\omega) = \mathcal{E}^{n-1}(\omega) - E_f\mathcal{E}^n(\omega)$, one sees that after multiplication by $\omega$ the matrices defined in Eq.~(\ref{eq:defVE_D}) simply need to be replaced as
\begin{equation}
\label{eq:powerw_Vw}
    V^{\omega}_{ij} \rightarrow V^{\prime \omega}_{ij} = V^\omega_{i(j-1)} + \delta_{0j} V_{i1}^{E},
\end{equation}
\begin{equation}
\label{eq:powerw_VE}
    V^{E}_{ij} \rightarrow V^{\prime E}_{ij} = V^{E}_{i(j+1)} - E_f V_{ij}^E.
\end{equation}
One repeats these transformations for each power of $\omega$ in the polynomial approximation of the flux.

We note that the simple polynomial approximation is not optimal, since it introduces higher-order moments in the expansion, which are more difficult to obtain from the Euclidean response than the $E_f$-dependent integrals.
Approximations of the flux in terms of inverse powers of $\omega$, e.g. ${1}/{(a+\omega)^n}$, are more appealing since these remove moments from the expansion.
Such dependence is only possibly suitable to describe the falling high-energy behavior of typical neutrino fluxes however.

\subsection{Fixed incoming energy}
We now consider the cross section for fixed $E$ and $q$ integrated over lepton energy
\begin{align}
    &\frac{\mathrm{d}\sigma(E)}{\mathrm{d}q} = G_F^2 q \nonumber \\
    &\times\int^{\omega_{max}}_0 \mathrm{d}\omega \frac{E - \omega}{E} \sum_i v_i(E,E-\omega,q) R_i(\omega,q).
    \label{eq:CS_diff_q_bound}
\end{align}
Again in the relativistic (massless) lepton limit.
The bounds on $q$ and $\omega$ that determine the physical region for scattering are set by energy conservation and
\begin{equation}
\label{eq:costheta_wE}
\cos\theta_f(E,E-\omega,q) = \frac{q^2 - E^2 - (E - \omega)^2}{-2(E-\omega)E}.
\end{equation}
For a fixed incoming energy $E$, the range of physical momentum transfers is given by $q\leq 2E$.
There are then two cases for $\omega_{max}$
\begin{equation}
    \omega_{\max} = 
    \begin{cases}
        q & \mathrm{if}~~~q < E \\
        2E - q & \mathrm{if}~~~ q > E
    \end{cases}~.
\end{equation}
The kinematic prefactors can be written as a linear combination of $\omega^n$, with at most $n=4$.
The expansion is given explicitly in Eqs.~(\ref{eq:VCC_DE_fixE}-\ref{eq:VTp_DE_fixE}).
Thus, if the upper limit of the integrals can be taken arbitrarily large, the single differential cross section of Eq.~(\ref{eq:CS_diff_q_bound}) is determined completely by the moments of the response given by Eq.~(\ref{eq:Int_Moments}).

In the case of $q > E$ the integral of Eq.~(\ref{eq:CS_diff_q_bound}) has to be cut-off at increasingly smaller values $\omega_{max} = 2E-q < q$. 
The case $q<E$ will be considered in the following. It is the more interesting case since one can consider the large $E$-limit at fixed $q$ and it includes only forward scattering angles, where the cross section typically peaks at large $E$.

\section{Numerical results for toy-model response}
\label{sec:toymodel}
We study the contributions of the integrals in Eqs.(\ref{eq:Int_Moments} - \ref{eq:intE2}) to the integrated cross sections using a toy-model nuclear response.
The $\omega$-dependence of the response is taken to be a Gaussian
\begin{equation}
R_{CC}\left(\omega,q\right) = \frac{1}{\sqrt{2\pi\sigma^2(q)}} \exp\left[ -\frac{\left( \omega - \mu(q) \right)^2}{2\sigma^2(q)} \right].
\end{equation}
For the peak and width of the response we use
\begin{equation}
    \mu(q) = \omega_{QE} = \frac{Q^2_{QE}}{2M} = \sqrt{q^2 + M^2} - M,
\end{equation}
and 
\begin{equation}
    3\sigma(q) = \sqrt{(q+k_F)^2 + M^2} - \sqrt{q^2 + M^2},
\end{equation}
where $M$ is the nucleon mass.
This is a semi-realistic proxy for the responses in the quasielastic region, which is the dominant reaction mechanism in the $q$-range of interest.
We set $R_{CL} = \frac{\omega}{q} R_{CC}$ and $R_{LL}= \frac{\omega^2}{q^2} R_{LL}$, i.e. we assume a conserved current.
For the transverse response we use $R_{T} = R_{T^\prime} = 1.2R_{CC}$, to include transverse enhancement.

Crucially, the Gaussian response decays before the upper bound of the integrals in Eqs.~(\ref{eq:CS_FF_Eq}) and (\ref{eq:CS_diff_q_bound}), that is $\omega_{max} = q$. 
Since there is negligible strength in the unphysical phase space we can take the upper bound of the integrals arbitrarily large. 
The main goal of this section is to study the relative importance of the different contributions to the observables under controlled conditions.
In section~\ref{sec:SF_integrals} we study realistic nuclear responses to address the feasibility of taking the upper limit of the integrals to infinity, and the influence of numerical noise.

\subsection{Fixed incoming energy}
\label{sec:Fixed_E_counting}
\begin{figure*}
    \centering
    \includegraphics[width=0.45\textwidth]{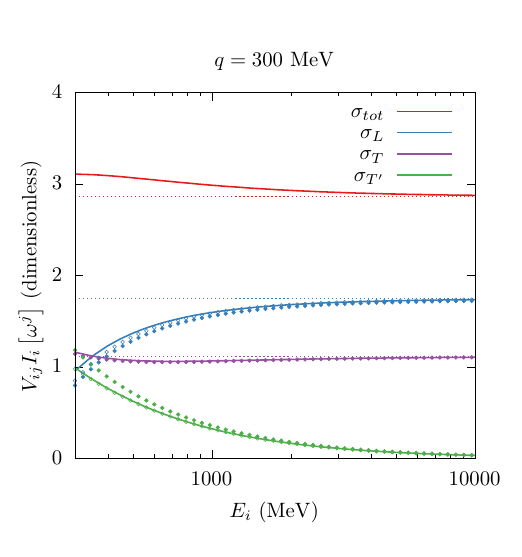}
    \includegraphics[width=0.45\textwidth]{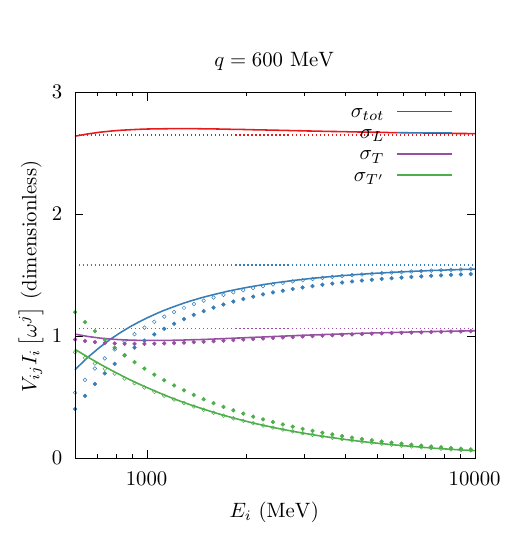} \\
    \includegraphics[width=0.45\textwidth]{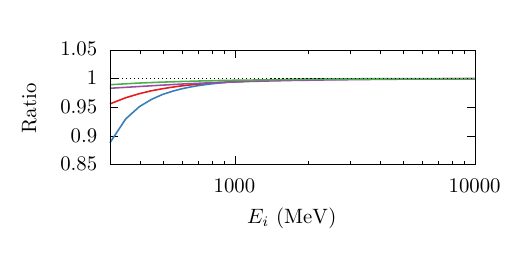}
    \includegraphics[width=0.45\textwidth]{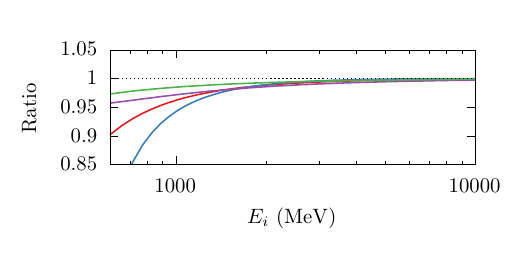}
    \caption{Longitudinal ($L$), transverse ($T$), and transverse-interference ($T^\prime$) contributions to the single-differential cross section at fixed $q=300$ MeV (left), and $q=600$ MeV (right). Solid lines show the exact result. Dashed lines are the high-energy limits. Filled points show the first energy-dependent contribution. Open symbols show the next order contribution, equivalent to keeping only contributions up to and including the second moments. The bottom panels show the ratio of this approximation to the full results.}
    \label{fig:CS_fixE_contributions}
\end{figure*}
The cross section integrated over the outgoing lepton energy at fixed $q$ and $E$ can be written as a linear combination of the $j$-th moment of the $i$-th response $I_i\left[ \omega^j  \right]$ as
\begin{equation}
    \label{eq:sigma_E_matrix}
    \frac{\mathrm{d}\sigma(E)}{\mathrm{d}q} = G^2q V^{\omega}_{ij} I_i\left[ \omega^j \right].
\end{equation}
The coefficients of the Matrix $V^{\omega}$ can be read off from Eqs.~(\ref{eq:VCC_DE_fixE}-\ref{eq:VTp_DE_fixE}), and involve up to the $4$-th moment.
The energy-dependence of the cross section is given by powers $1/{E^n}$ with $n \in \{0,1,2\}$.

It is straightforward to organize the contributions to the cross section in terms of ratios of energy scales.
The response in the quasielastic region is characterized by the energy scale
\begin{equation}
    \mu \sim \sigma \sim \frac{q^2}{2M} + O(q^3/M^2),
\end{equation}
which follows from the non-relativistic expansion of the peak position\footnote{In the relativistic case $\mu \sim \omega_{QE} = \frac{Q^2_{QE}}{2M}$, and $\mu^n/q^n \sim (\frac{\omega_{QE}}{q})^n < 1$ is still a suitable small parameter.}. The relevant energy scales are then $q, E$, and $M$.
The moments may be organized as follows
\begin{equation}
I_{CC}\left[ w^0 \right] \sim I_{T}\left[ \omega^0\right] \sim I_{T^\prime}\left[ \omega^0 \right] \sim 1,
\end{equation}
which is simply the normalization of the response.
The higher order moments scale as
\begin{equation}
    \frac{I_{i}\left[ w^n \right]}{q^n} \sim \frac{\mu^n}{q^n} \sim \left(\frac{q}{2M}\right)^n,
\end{equation}
for $i \in \{CC,T,T^\prime\}$.
Since we are assuming a conserved current, the $CL$ and $LL$ responses are suppressed by additional powers\footnote{This additional suppression for the $CL$ and $LL$ responses should be modified when non-conserved currents are considered, e.g. the axial current. To be conservative one may simply count them in the same way as the other responses.}
\begin{equation}
        \frac{I_{CC}\left[ \omega^n \right]}{q^n} = \frac{I_{CL}\left[\omega^{n-1} \right]}{q^{n-1}} = \frac{I_{LL}\left[\omega^{n-2} \right]}{q^{n-2}}. 
\end{equation}
In the low energy region where $E \sim q$, these assignments give the relative importance of the contribution from each moment.

In the high energy limit, $E >> 2M_N$, the cross section is given by
\begin{align}
    \frac{\mathrm{d}\sigma(E\rightarrow \infty)}{\mathrm{d}q} &= 2 M_{CC}^0 - \frac{4}{q} M_{CL}^1 + \frac{2}{q^2}M_{LL}^2 \nonumber\\
    & + M_{T}^0 -\frac{1}{q^2}M_T^2,
    \label{eq:sigma_E_HE}
\end{align}
where we use the shorthand $M^{n}_i \equiv I_i\left[\omega^n \right]$ for the moments.
The high-$E$ limit involves at most the second moments of responses, which contribute at order $(q/2M)^2$ in the transverse and $(q/2M)^4$ in the longitudinal case.

In the intermediate energy region, where $E \sim 2 M_N$, one has the following hierarchy
\begin{align}
    &\frac{\mathrm{d}\sigma(E_i \sim 2M_N)}{\mathrm{d}q} = 2M_{CC}^0 + M_{T}^0  \label{eq:M_Efix_0}\\
    &+ \frac{q}{E_i} M_{T^\prime}^0 \label{eq:M_Efix_1}\\
    &-\frac{2}{E_i} M_{CC}^1 - \frac{q^2}{2E_i^2} M_{CC}^0  - \frac{4}{q} M_{CL}^1   \nonumber \\
    &-\frac{1}{E_i}M_{T}^1 + \frac{q^2}{4E_i^2} M_T^{0} - \frac{1}{q^2}M_{T}^2 \label{eq:M_Efix_2} \\
    & - \frac{q}{2E_i^2} M_{T^\prime}^1 - \frac{1}{qE_i} M_{T^\prime}^2 + O(\frac{q^4}{M^4}) \label{eq:M_Efix_3}
\end{align}
where each numbered line corresponds to contributions of increasing order up to and including $(q/2M)^3$.

The relative contributions of the moments to the cross section are shown in Fig.~\ref{fig:CS_fixE_contributions} for $q=300$ and $q=600$ MeV.
We show the partial cross sections split up into longitudinal ($CC+CL+LL$), transverse ($T$) and transverse-interference ($T^\prime$).
The first energy-dependent contribution for the transverse cross section, that is the $M_T^{n}$ terms in Eq.~(\ref{eq:M_Efix_2}), i.e. up to order $(q/2M)^2$, is enough to reproduce the transverse cross section with accuracy better than $5\%$ over the whole energy range. This includes all the transverse contributions to the high-energy limit of Eq.~(\ref{eq:sigma_E_HE}).
For the $L$ and $T^\prime$ cross sections, the next order contributions are required. These are Eq.~(\ref{eq:M_Efix_3}) for $T^\prime$ and the $(q/M)^4$ contribution (not given explicitly) for $L$.
The $M_{LL}^2$ contribution to the high-energy limit comes at this order.
This level of approximation corresponds to including all contributions up to and including the second moment. 
In this case one reproduces the total cross section over the whole energy range with an accuracy of better than $5\%$ for $q=300$ and $10\%$ for $q=600$. The deviation is mostly due to the longitudinal response.

\subsection{Unit flux}
\begin{figure*}
\includegraphics[width=0.45\textwidth]{"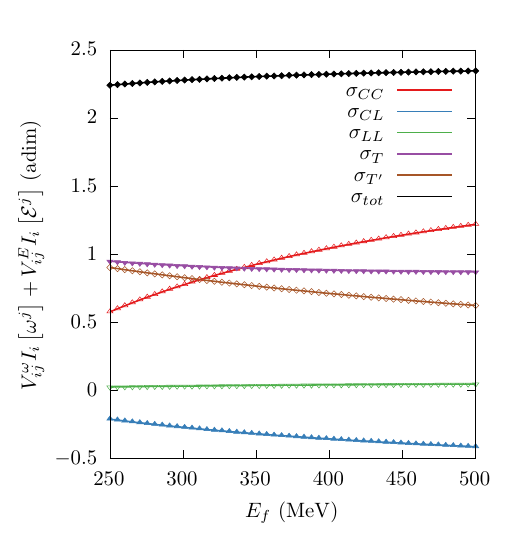"}
\includegraphics[width=0.45\textwidth]{"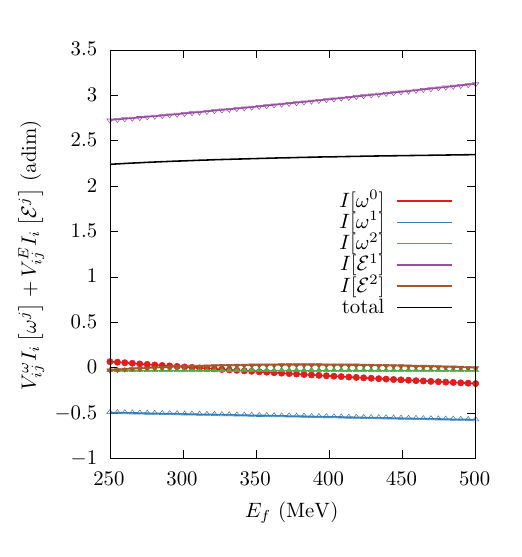"}
\caption{Contributions to the energy-integrated cross section, obtained in $\omega$ space (lines), and directly from the Euclidean response (symbols). The left panel shows the cross section decomposed by contribution of the different responses. The right panel shows the contribution from terms proportional to integrals of responses weighted with $\omega^n$ and $\mathcal{E}_n$.}
\label{fig:comp_lap}
\end{figure*}
We now consider the energy-averaged cross section. As described in Sec.~\ref{sec:unitflux_bounds}, the cross section can be written in terms of the integrals of Eqs.~(\ref{eq:Int_Moments}-\ref{eq:intE2}). 
We compute these contributions from the Euclidean responses obtained with the Gaussian toy-model.
The sum rule $I_i\left[\omega^0\right] = E_i(0)$ and the $E_f$-dependent integrals are  straightforward to obtain from the Euclidean responses.
The only non-trivial integrals are the first and second moments $I_i\left[ \omega^1\right]$ and $I_i\left[ \omega^2 \right]$. 
These could be obtained by evaluating the derivatives of the Euclidean response at $\tau=0$.
We use instead the method laid out in Appendix~\ref{app:Moment_calculation} to compute them.
For the purpose of this section we don't consider the numerical uncertainty. A realistic numerical uncertainty is considered in Sec.~\ref{sec:SF_integrals}.

Results for the energy-integrated cross section as function of $E_f$ for $q=500~\mathrm{MeV}$ are shown in Fig.~\ref{fig:comp_lap}.
The left panel shows the cross section separated by different response functions. 
The agreement between results obtained in the $\omega$ and $\tau$ domain is excellent, which is expected since there is negligible strength outside of the physical region for these responses. 
The right-hand panel shows the result split up in terms of the contributions proportional to the different $\omega$-dependencies. 
While all responses except for $\sigma_{LL}$ provide non-negligible contributions to the total rate, the integrals  $I_i\left[\omega^1\right]$ and $I_i\left[\mathcal{E}^1\right]$ provide by far the largest contribution to the cross section. 
The fact that the contribution from the second moment is negligible, is a robust result. An $\omega^2$ dependence only enters in the $LL$ and $T$ contributions, see Eqs.~(\ref{eq:VCC_DE}-\ref{eq:VTp_DE}), contributing as $(q/2M)^4$ and $(q/2M)^2$ respectively. Additionally, a small cancellation occurs between these contributions which have opposite sign.
That the sum rule and the integrals over $\mathcal{E}^2$ give such small contributions requires cancellations between the different partial cross sections, and might not hold for a different model of the responses. 

\subsection{Flux-averaged cross section}
\begin{figure}
    \centering
    \includegraphics[width=0.5\textwidth]{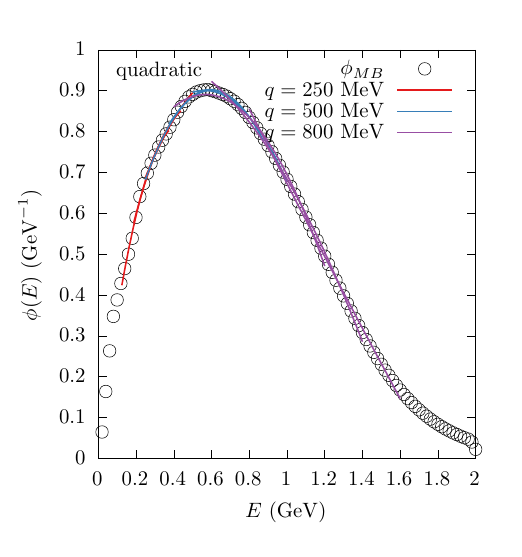}
    \caption{Fits of the MiniBooNE flux with second order polynomials in windows of energy $ E\in [E_0,E_0+q]$ for different values of $q$. For each $q$-value three $E_0$ windows are shown $E_0 \in \{q/2, 3q/4, q\}$, equally spaced in the physical region $E\in [q/2,q]$.}
    \label{fig:flux_fit}
\end{figure}
We now extend the previous result to a realistic neutrino flux, we use the MiniBooNE flux~\cite{MBflux:2009} in the following. 
We fit the flux with second order polynomials in $E$ in each of the energy windows with width $q$ accessible at fixed $E_f$ as described in section~\ref{sec:fluxfolding}. 
Fit results are shown for a couple of values of $q$ and windows of energy in Fig~\ref{fig:flux_fit}. 
Of course one can consider different functional forms, even for each of the energy windows.
\begin{figure*}
    \includegraphics[width=0.45\textwidth]{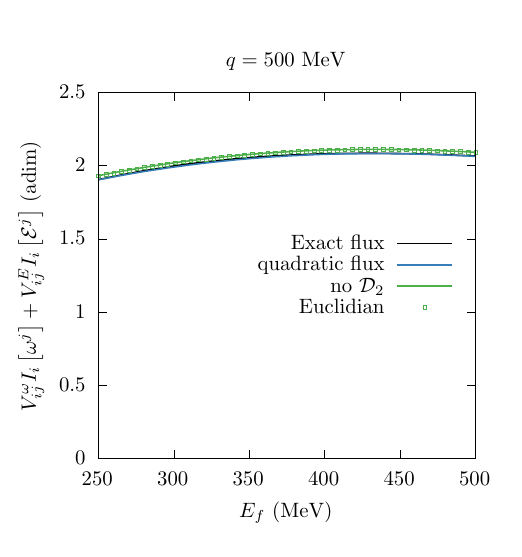}
    \includegraphics[width=0.45\textwidth]{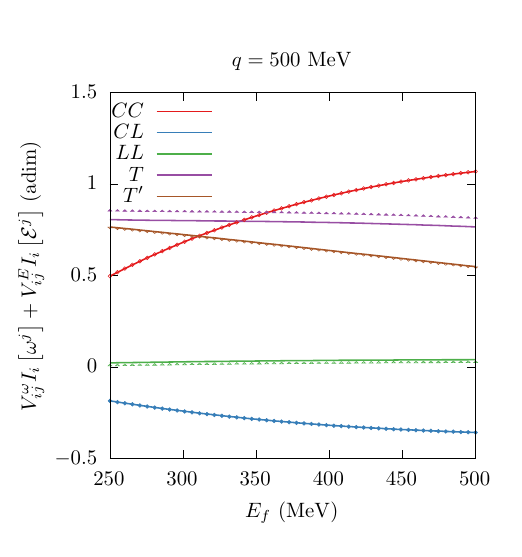}
    \caption{MiniBooNE flux averaged cross sections. Symbols are obtained from the Euclidean response, lines are calculated from the responses in $\omega$-space. The left-hand panel shows the total cross section. Black line is the result with the exact MiniBooNE flux, the blue line is the quadratic approximation to the flux. The green line is the calculation with the quadratic flux where additionally the contribution of $\omega^2$ terms is dropped in the lepton prefactors. Green open symbols are the equivalent calculation directly from the Euclidean response. The right-hand panel shows the exact results for the different responses, compared to the result from the Euclidean response. The difference is due to dropping the $\omega^2$ term in the latter.}
    \label{fig:MB_FF_q500}
\end{figure*}
The downside of a polynomial fit is of course that higher-order moments of the response need to be computed, while in the previous section the first and second moment sufficed.
We compute the third moment with the same method described in appendix~\ref{app:Moment_calculation}.
In principle we also require the fourth moment $I_i\left[ \mathcal{\omega}^4\right]$ for the $LL$ and $T$ responses. 
However, as we have shown in the previous section (see Fig.~\ref{fig:comp_lap}), the contribution to the energy-averaged cross section proportional to the second moments is highly suppressed. 
It is hence feasible to drop these contributions.
Therefore, in the matrix of Eq.~(\ref{eq:defVE_D}), we set $V_{i2}^{\omega} = 0$, i.e. neglecting the $\omega^2$ terms in Eqs.~(\ref{eq:VCC_DE} - \ref{eq:VTp_DE}).
The linear combination of integrals that describes the flux averaged cross section are obtained by applying the transformations of Eq.~(\ref{eq:powerw_Vw}-\ref{eq:powerw_VE}) for each power of $\omega$ in the polynomial fit.
The highest moment that needs to be computed for the flux-averaged cross section is then the third moment. 

Results for the MiniBooNE flux-averaged cross section are shown in Fig.~\ref{fig:MB_FF_q500}.
The left panel shows the exact calculations done in $\omega$-space by solid lines. We compare the calculation using the exact MiniBooNE flux to the one obtained with the quadratic approximation, agreement is excellent.
The green line shows the effect of discarding the contribution of the second moment (using also the quadratic approximation to the flux). 
The latter can be compared to the calculation based on the Euclidean response, where we also discard these contributions.
It is seen that for the overall flux-averaged cross section, this leads to a smaller than $5\%$ discrepancy with the full result.

The right-hand panel of Fig.~\ref{fig:MB_FF_q500} shows the breakdown into different responses, where lines denote the exact calculations (with quadratic approximation of the flux) done in $\omega$-space and corresponding symbols the results obtained from the Euclidean response. 
The differences are solely due to dropping the $\omega^2$ terms, which only occur in the $T$ and $LL$ contributions. 
Clearly the strength missing in the total cross section stems practically completely from the transverse contribution.

\section{Realistic nuclear spectral function}
\label{sec:SF_integrals}
In the Gaussian toy model the strength of the response for $\omega > q$ is negligible.
For a realistic nuclear response, there will be non-negligible contributions in the unphysical region. 
Moreover, in the previous section we didn't consider numerical uncertainty.
We study the effect of these complications on the computation of the integrals in Eqs.~(\ref{eq:Int_Moments}-\ref{eq:intE2}) by using a realistic model for the response and for the numerical uncertainty.

We use Euclidean responses obtained in the PWIA using the spectral function for ${}^{12}$C obtained from Variational Monte Carlo (VMC) calculations as discussed in Ref.~\cite{Lovato:2023khk}.
The nuclear responses are 
\begin{align}
    R_i(\omega,q) &\sim \int\mathrm{d}\mathbf{k}\mathrm{d}E~ \frac{M_N}{E_N} S(E, k) h_i^{s.n.} \nonumber \\
    &\times \delta(\omega - E + M_N- E_N(\mathbf{q},\mathbf{k})) \nonumber \\
    &= 2\pi \frac{M_N}{q} \int_{0}^{\omega} \mathrm{d}E \int_{k^-(\omega,q,E)}^{k^{+}(\omega,q, E)} k\mathrm{d}k~h_i^{s.n} S(E, \lvert \mathbf{k} \rvert).
    \label{eq:response_PWIA}
\end{align}
where $h_i^{s.n.}$ is a single-nucleon response, and $S(E,\mathbf{k})$ is the nuclear spectral function. 
The missing energy and momentum are denoted $E,\mathbf{k}$ and the outgoing nucleon kinematics are treated relativistically $E_N(\mathbf{q},\mathbf{k}) = \sqrt{(\mathbf{q}+\mathbf{k})^2 + M_N^2}$.
The bounds in the momentum integral are discussed in appendix~\ref{app:highK_SF}.

The spectral function used in this work is described in detail in Refs.~\cite{CLAS:2022odn,Lovato:2023khk}. Using the realistic AV18 nuclear potential, which features a strong short-range repulsive core, leads to pronounced high-momentum tails in nucleon momentum distributions, reflecting the presence of short-range correlated nucleon pairs.
The high-$\omega$ tail of the response, which is the main contribution in the unphysical region, is determined by the high-$\lvert \mathbf{k} \vert$ behavior of spectral function since $\omega \sim T_N = \sqrt{(\mathbf{q} + \mathbf{k})^2 + M_N^2}$.
The high-$\mathbf{k}$ behavior of the single-N (neutron or proton) spectral function is determined by the two-nucleon momentum distribution obtained from VMC calculations~\cite{nkk_web},
\begin{align}
    S_{N}(E,\mathbf{k}) = &\sum_{\tau_{k^\prime} = {p,n}} \mathcal{N}_{\tau_{k\prime}}\int \frac{\mathrm{d}\mathbf{k}^\prime}{(2\pi)^3} \nonumber \\
    &\times n_{N,\tau_{k^\prime}}(\mathbf{k}, \mathbf{k}^\prime) \delta\left(E - E_c - T_{A-1}(\mathbf{k},\mathbf{k}^\prime) \right).
    \label{eq:SF_VMC_unfactorized}
\end{align}
Here the integral is over the momentum of the second nucleon $\mathbf{k}^\prime$ and $n_{\tau,\tau^\prime}(\mathbf{k},\mathbf{k}^\prime)$ is the two-nucleon momentum distribution for nucleon pairs with isospins $(\tau, \tau^\prime) = (n,n),~(n,p)~\mathrm{or}~(p,p)$.
The kinetic energy of the $A-1$ system is comprised of the kinetic energy of the second nucleon and the residual $A-2$ system $T_{A-1}(\mathbf{k},\mathbf{k}^\prime) = \mathbf{k^\prime}^2/(2M) + \left(\mathbf{k} + \mathbf{k}^\prime\right)^2/(2M_{A-2})$ and $E_c$ is a fixed two-nucleon separation energy.

We compute the Euclidean response as the Laplace transform of Eq.~(\ref{eq:response_PWIA}), and use it to compute Eqs.~(\ref{eq:Int_Moments}-\ref{eq:intE2}).
Note that we fix the value of the nucleon form factors that enter in the single-nucleon responses to their values at the quasielastic peak, determined by $\omega_{QE}= \sqrt{q^2 + M_N^2} -M_N$.
This is also the case in GFMC calculations, and is discussed in section~\ref{sec:ormFactors}.

We consider two sources of numerical uncertainty.
First the Euclidean responses are discretized on a grid with fixed stepsize $\Delta \tau$.
Second, we include an uncertainty on the value of the Euclidean response.
For this we assume that we have $N$ samples of the Euclidean response at each gridpoint. The variance of the samples is given by a simple model
\begin{equation}
    \sqrt{Var\left[ E(\tau)\right]} = f_0 E(0) \sqrt{ \frac{E(0)}{E(\tau)}},
\end{equation}
where $f_0$ is the relative variance assigned to $E(0)$.
This model for the variance provides an excellent reproduction of the variance computed from samples of the Euclidean responses obtained with GFMC in Ref.~\cite{PhysRevX.10.031068}. 
To propagate the uncertainty to the observables considered, we compute the average and variance of observables obtained from $N$ samples of the Euclidean response.
Each of the samples is obtained as  
\begin{equation}
E_j(i\Delta\tau) = E(i\Delta\tau)\left(1 + x_{j,i}f_0 \left(\frac{E(0)}{E(i\Delta\tau)}\right)^{3/2} \right),
\end{equation}
where $i\Delta\tau$ is the $i$-th gridpoint and $j \in \{1,\ldots,N\}$ the $j$-th sample.
$x_{j,i}$ is a random variable drawn from a Gaussian with unit variance and $E(\tau)$ the Euclidean response computed from the spectral function.
The stepsize $\Delta \tau$, relative uncertainty $f_0$, and the number of samples $N$ can be varied to study the numerical methods. 
For the GFMC responses of Ref.~\cite{PhysRevX.10.031068} one has $\Delta \tau = 10^{-3}~\mathrm{MeV}^{-1}$, $f_0 \approx 0.003 -0.01$ (depending on the response), and $N \sim 50 - 100$. 

\subsection{High-$\omega$ corrections}
\begin{figure}
    \centering
    \includegraphics[width=0.5\textwidth]{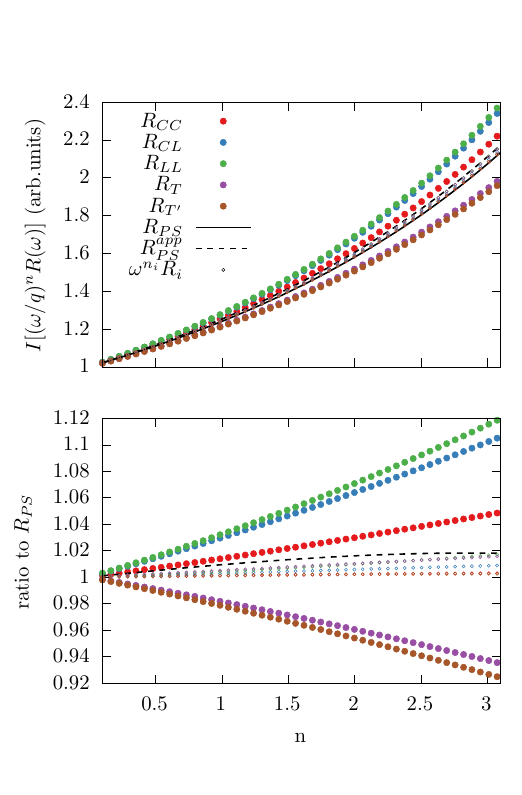}
    \caption{Contribution from the unphysical region to the $n$-th moment of the responses, Eq.~(\ref{eq:approx_omega_unphys}) for $q=500$. We scale all moments by $1/q^n$, and normalize to 1 for $n=0$.
    The filled circles denote the different nuclear responses, the solid line is the phase-space response $R_{PS}$ of Eq.~(\ref{eq:R_PS_SF}). The ratio is shown in the bottom panel. The empty circles are the results when the responses are scaled by appropriate powers of $\omega$ from Eq.~(\ref{eq:R_w_scaling}). The dashed line is the approximation for $R_{PS}$ from Eq.~(\ref{eq:R_app_ndist}) determined from the momentum distribution.}
    \label{fig:R_w_q500_unphys}
\end{figure}
Responses computed with the spectral function have non-negligible strength in the region $\omega > q$.
Hence, if one computes the integrals of Eqs.~(\ref{eq:Int_Moments}-\ref{eq:intE2}) from the Euclidean responses there will be an unphysical contribution to the integrals of Eqs.~(\ref{eq:CS_unitflux}), and (\ref{eq:CS_diff_q_bound}). This contribution can be estimated and subtracted.
This is tractable since the response at large-$\omega$ is determined by high-momentum parts of the spectral function. 
This part of the spectral function is driven by the short-range nucleon-nucleon interaction and leads to approximately universal high-momentum dependence in different nuclei~\cite{Tropiano:2021qgf, Ryckebusch:2019oya, Cruz-Torres:2019fum, CiofidegliAtti:2015lcu, Ryckebusch:2014ann, CLAS:2022odn}.
To illustrate this we compute a simple correction for the strength in the unphysical region which is applied to all responses.
We define the response that includes the phase space integral but omits the single nucleon response
\begin{equation}
\label{eq:R_PS_SF}
    R_{PS}(\omega,q) = 2\pi \frac{M_N}{q} \int_{0}^{\omega} \mathrm{d}E \int_{k^-(\omega,q,E)}^{k^{+}(\omega,q, E)} k\mathrm{d}k~S(E, k).
\end{equation}
If the $\omega$-dependence of this response is a fair approximation to that of the actual responses in the unphysical region, the relevant integrals of the responses should be well approximated by those of $R_{PS}$
\begin{equation}
    \label{eq:approx_omega_unphys}
    \int_q^\infty \mathrm{d}\omega~\omega^n R_i(\omega) \approx f_{i} \int_q^{\infty}\mathrm{d}\omega~\omega^n R_{PS}(\omega),
\end{equation}
and,
\begin{equation}
\label{eq:approx_Epsilon_unphys}
    \int_q^\infty \mathrm{d}\omega \frac{1}{(\omega + E_f)^n} R_i(\omega) \approx f_{i} \int_q^{\infty}\mathrm{d}\omega~\mathcal{E}^n R_{PS}(\omega).
\end{equation}
Here $i \in \{CC,CL,LL,T,T^\prime\}$ and $f_i$ is a normalization factor that depends on the response.
In the PWIA it is determined by the single nucleon response $h_i^{s.n.}$.
For responses computed in GFMC one needs to estimate these normalization factors, e.g. by using the PWIA.
We here compute the normalization factors as the fraction of total strength in the unphysical region for each of the responses
\begin{equation}
\label{eq:contacts}
    f_i = \frac{\int_q^{\infty} R_i(\omega)~\mathrm{d\omega}}{\int_0^{\infty} R_i(\omega)~\mathrm{d}\omega }.
\end{equation}
The ratios are found to be $f_{CC} \approx 0.04,~f_{CL} \approx 0.06,~ f_{LL} \approx 0.03,~f_{T} \approx f_{T^\prime} \approx 0.015$.
Note that, if indeed the shape of the high-$\omega$ response is somewhat universal, one might be able to deduce the relative size of these normalization factors from the integrals computed from the Euclidean response which includes the unphysical region.

It is clear that the strength in the unphysical region from the energy-dependent integrals of Eq.~(\ref{eq:approx_Epsilon_unphys}) is strongly suppressed compared to those of the moments.
In Fig.~\ref{fig:R_w_q500_unphys} we show the quality of the approximation of Eq.~(\ref{eq:approx_omega_unphys}) for the moments.
The relative error grows approximately linearly with $n$. We overestimate the transverse responses while underestimating the longitudinal responses.
The deviations from a universal $\omega$-dependence of the response at large energy transfer is found to be approximately given by small powers of $\omega$
\begin{equation}
\label{eq:R_w_scaling}
    \omega R_{PS} \sim \omega^{1/2} R_{CC} \sim R_{CL} \sim R_{LL} \sim \omega^2 R_{T} \sim \omega^2 R_{T^\prime}.
\end{equation}
Taking this scaling into account decreases the error, it is at most $2\%$ for $n=3$ as shown in Fig.~\ref{fig:R_w_q500_unphys}.
Of course such corrections depend on the nuclear current.
In the following we don't take these additional corrections into account, we use the integrals computed from $R_{PS}$ to determine a universal $n$-dependence of the strength in the unphysical region.
Keep in mind that this is the error on a correction, not the error on the actual moments that enter calculations of the cross section.

Lastly, $R_{PS}$ at high-$\omega$ is determined by nucleons at high missing momentum.
As discussed in appendix~\ref{app:highK_SF}, the energy and momentum distribution, along with the response $R_{PS}$ are then almost completely determined by the two-nucleon relative momentum distribution.
An approximation for $R_{PS}$ may be obtained in this case from the single-nucleon momentum distribution $n(k)$ instead of the spectral function\footnote{This integral in more familiar form is given in  Eq.~(\ref{eq:R_PS_scaling}).}
\begin{align}
\label{eq:R_app_ndist}
    R_{PS}^{app}(\omega,q) = 2\pi \frac{M^2}{q}\int_{T^{-}(\omega,q)}^{T^{+}(\omega,q)} \mathrm{d}T~n\left( \sqrt{2M T} \right).
\end{align} 
This approximation may be derived by neglecting center-of-mass motion of nucleon pairs for a spectral function of the form of Eq.~(\ref{eq:SEk_fact_CDA}).
The bounds on the integral are determined by Eq.~(\ref{eq:ineq_bounds}).
Corrections for center-of-mass motion are discussed in some detail in appendix~\ref{app:highK_SF}.
Note that we restrict the upper bound to $T^+(\omega,q) < 300~\mathrm{MeV}$ here, this is the cut-off in missing energy used to compute all responses.
The result is shown by the dashed line in the upper panel of Fig.~\ref{fig:R_w_q500_unphys}.
Clearly this approximation is excellent, showing that corrections for the unphysical region may be obtained from the nucleon momentum distribution.

\subsection{Results for the integrals}
\begin{figure}
    \centering
    \includegraphics[width=0.45\textwidth]{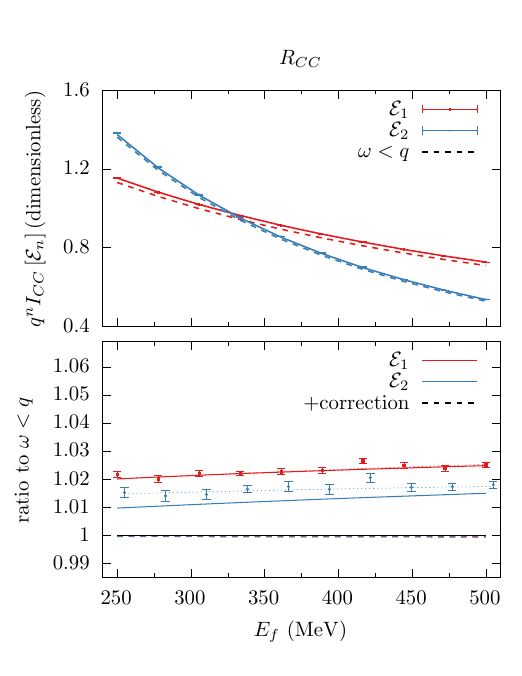}
    \caption{Results of the integrals $I_{CC}\left[ \mathcal{E}^n\right]$ obtained from the Euclidean response. Solid lines show the exact result. Dotted lines are the result obtained on a grid with fixed stepsize. The errorbars show the uncertainty from the variance of the Euclidean response. The dashed lines in the top panels show the exact result restricted to the physical region $\omega < q$. The bottom panel shows the ratio to the result in the physical region. The dashed lines in the bottom panel are obtained after subtracting the correction for the high-$\omega$ tail of the response. }
    \label{fig:R_CC_E_ratio}
\end{figure}
\begin{figure}
    \centering
    \includegraphics[width=0.45\textwidth]{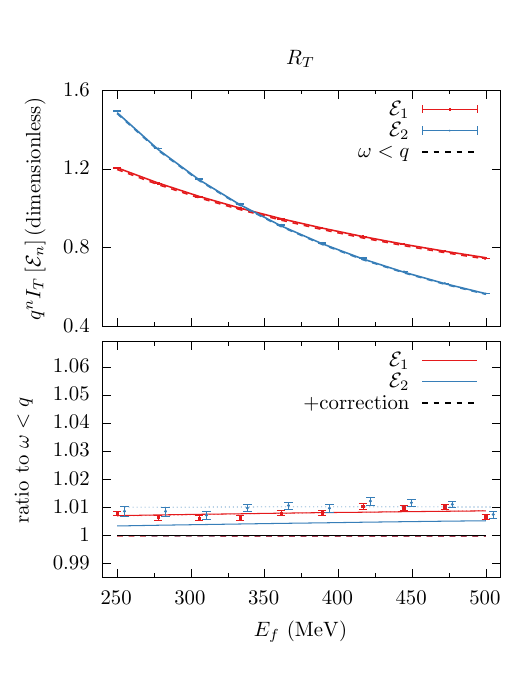}
    \caption{Same as in Fig.~\ref{fig:R_CC_E_ratio} but for the transverse response.}
    \label{fig:R_T_E_ratio}
\end{figure}
The results for the integrals over the response weighted with $\frac{1}{(E + \omega)^n}$ for $R_{CC}$ and $R_{T}$ are shown in Figs.~\ref{fig:R_CC_E_ratio}~and~\ref{fig:R_T_E_ratio} respectively. Results for the other responses are similar and collected in Appendix~\ref{app:Eints}.
For ease of comparison, the responses are normalized so that $E(0) = 1$.
We present dimensionless results by weighing with a factor $q^n$. 

The integrals have the form $\int f(\tau) e^{-\Lambda\tau} \mathrm{d}\tau$.
We compute these on a grid of fixed stepsize $\Delta \tau = 10^{-3}~\mathrm{MeV}^{-1}$, by approximating $f(\tau)$ as linear between gridpoints. 
To improve convergence of the integrals we divide out the Euclidean response for a free nucleon, $E_{free}(\tau,q) = e^{-\tau~\omega_{QE}}$, where $\omega_{QE}(q)= \sqrt{q^2 + M_N^2} - M_N$. I.e. the integrals are computed by treating $f(\tau) \equiv \tau^{n-1}E(\tau)/E_{free}(\tau,q)$ as linear between gridpoints while the exponential decay with $\Lambda = E + Q$ is included exactly. 
With this approach, the numerical uncertainty due to the gridsize is negligible for $n=1$, and sub-percent for $n=2$ (dotted versus solid lines in the figures).
A trapezoid rule integral with the same gridsize leads to $1\%-5\%$ discrepancies depending on the response. The accuracy in the $n=2$ case can be improved straightforwardly by treating the factor $\tau$ exactly.
The uncertainty due to numerical noise, here obtained with conservative values, from $N=50$ samples with $f_0 = 0.01$, is found to be sub-percent level. 

The contribution from the unphysical phase space to these integrals is $2-3$ percent for $n=1$ and $1-2$ percent for $n=2$. 
The contribution is smaller for the transverse responses, which is understood from Eq.~(\ref{eq:R_w_scaling}).
One sees that for each of the responses, the correction obtained from the phase space response of Eq.~(\ref{eq:approx_Epsilon_unphys}) provides excellent results. 
With this correction, the result restricted to the physical region is reproduced at permille level. 

\begin{figure}
    \centering
    \includegraphics[width=0.45\textwidth]{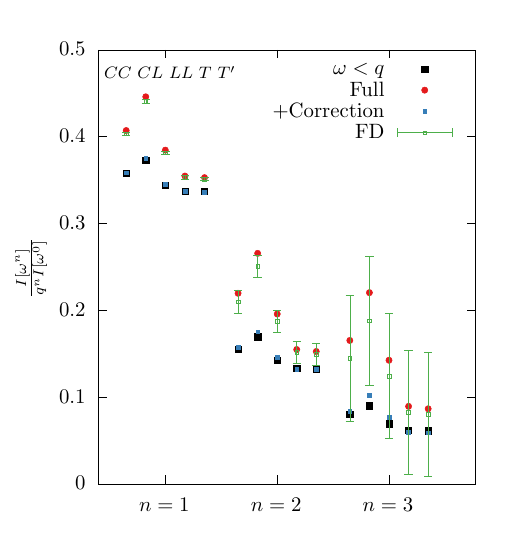}
    \caption{The first, second and third moments of the five nuclear responses. Solid squares show the results restricted to the physical region $\omega < q$ which enter in the neutrino cross section. Red circles show the moments of the full response. Blue squares show the result for the moments after applying the correction determined from the nuclear momentum distribution. Finally, numerical results obtained with forward-difference on a fixed grid are shown by green open circles. The errorbars correspond to the uncertainty obtained from $75$ samples of the Euclidean response and $f_0=0.003$. }
    \label{fig:Moments_corr_numerical}
\end{figure}
The results for the moments are shown in Fig.~\ref{fig:Moments_corr_numerical}.
We present the dimensionless fractions $I_i\left[ \omega^n \right]/q^n$. When the integrals are restricted to the physical region, these approximately behave as $(\omega_{QE}/q)^n$, as in Sec.~\ref{sec:Fixed_E_counting}.
The excess strength in the unphysical region ranges from $\sim 5\%$ for the first moment of the transverse responses to $\sim 100 \%$ for the third moment of the longitudinal responses. 

If we subtract the correction for the unphysical region from the full moments we find excellent results for the first and second moments. Only the third moment of the longitudinal responses is still overpredicted by around $10\%$. In this case the size of the correction is of the same size as the moment, and given the simplicity of the correction factor the agreement even in this case is quite remarkable.

We now turn to the numerical calculation of the moments from the Euclidean response when a realistic numerical uncertainty is taken into account.
The method described in Appendix~\ref{app:Moment_calculation} is only suitable if $\Delta\tau \times \Lambda \ll 1$ when integrals are computed on a grid with fixed stepsize $\Delta\tau$, otherwise it is equivalent to finite-difference\footnote{This is easily verified numerically, but can be illustrated by a simple analytical example: the Euclidean response for a free nucleon $E(\tau) = e^{-\tau Q}$. In this case Eq.~(\ref{eq:omega_cut}) is approximated on a grid as $I\approx\frac{1 - e^{-Q\Delta\tau }}{\Delta\tau} + \sum_{i=1} e^{-(\Lambda + Q)\Delta\tau i}\frac{2}{\Delta \tau}\left[\cosh(\Delta\tau~Q) - 1 \right]$. When $\Delta\tau \ll Q + \Lambda$, summing the geometric series yields $I \approx -Q + \frac{Q^2}{\Lambda} -\frac{Q^3}{\Lambda^2} + \ldots $. If $\Delta\tau\times\Lambda > 1$, the exponential decay dominates and we get $I \approx \frac{1 - e^{-Q\Delta\tau }}{\Delta\tau}$, which is the first order forward difference formula.}. 
Hence, we use the forward difference formulae to compute the moments from the Euclidean response. 
We again divide out the $\tau$-dependence of the free-nucleon response using finite difference on $f(\tau)$ defined as $E(\tau) \equiv f(\tau) e^{-\omega_{QE}} $ to determine the derivatives of $E(\tau)$.
The finite-difference results shown in Fig.~\ref{fig:Moments_corr_numerical} are the exact values obtained with a stepsize $\Delta\tau = 10^{-3}~\mathrm{MeV}^{-1}$. This is the value in the limit of infinite samples of the Euclidean response or negligible $f_0$. 
The error-bars show the 1-$\sigma$ uncertainty on this value obtained with $N=75$ samples and with $f_0 = 0.003$, the errors are of course proportional to $f_0$ and to $1/\sqrt{N}$.
The results for the first moment are computed with second order (3-point) forward difference, and reproduce the exact values at percent level. The uncertainty due to variance in samples of the Euclidean response is also percent-level in this case.
The second and third moment are underestimated more significantly. The second derivative of $f(\tau)$ is computed with 4-point forward difference in this case, while the third derivative is computed with 3-point forward difference.
The uncertainties can of course be reduced by increasing the sample size $N$. 
The limiting factor then becomes the gridsize, we used $\Delta \tau = 10^{-3}~\mathrm{MeV}^{-1}$, this value needs to be reduced to improve agreement with the exact result or a more robust method for computing the moments needs to be used.

\section{Some further considerations}
\label{sec:considerations}
We have expressed energy integrated neutrino scattering cross sections in the zero-lepton mass limit in terms of a limited set of integrals over the nuclear response weighted by functions of energy.
If the upper bound of these integrals can be taken arbitrarily large, which is the case if the responses are negligible outside of the accessible phase space, then one can compute the required integrals directly from the Euclidean response.
Robust corrections for strength in the unphysical region was computed in the PWIA with realistic spectral functions. In these calculations, as is the case in GFMC calculations of the Euclidean response, the nucleon form factors were fixed at their value near the quasielastic peak. 
In the following we address some of these assumptions.
\subsection{Massive leptons}
\label{sec:leptonmass}
We have given the necessary expressions in the relativistic limit, exact for massless outgoing leptons, suitable for charged-current interactions of electron neutrinos in the energy range considered.
For massive outgoing leptons, e.g. muon neutrinos at MeV energies, lepton mass corrections are required.

These corrections are most readily obtained from the expressions of Refs.~\cite{Moreno:2014kia} which explicitly separate lepton mass terms, we adapt their result here.
The cross section is given by
\begin{equation}
    \frac{\mathrm{d}\sigma(E)}{\mathrm{d}E_f\mathrm{d}q} = G_F^2 q \frac{v_0}{E^2} \sum_{i} \hat{V}_i R_i,
\end{equation}
where the lepton factors\footnote{We implicitly absorb a factor $-2$ in $\hat{V}_{CL}$ and $\hat{V}_{T^\prime}$ compared to those of Ref.~\cite{Moreno:2014kia}.} $\hat{V}_i$ are given in Ref.~\cite{Moreno:2014kia}, and
\begin{equation}
    v_0 \equiv (E+E^\prime)^2 - q^2.
\end{equation}
For fixed values of $E,E_f,q$ the required lepton factors are
\begin{align}
    \frac{v_0}{E^2} \hat{V}_{CC} &= \frac{E_f}{E}v_{CC}^{m=0} - \frac{m^2}{2E^2}, \\
    \frac{v_0}{E^2} \hat{V}_{CL} &= \frac{E_f}{E}v_{CL}^{m=0} - \frac{m^2}{E^2}\frac{E+E^\prime}{q}, \\
    \frac{v_0}{E^2} \hat{V}_{LL} &= \frac{E_f}{E}v_{LL}^{m=0} + \frac{m^2}{2E^2 q^2}\left[ q^2 + 2\omega(E+E^\prime) + m^2 \right], \\
    \frac{v_0}{E^2} \hat{V}_T &= \frac{E_f}{E} v_T^{m=0} - \frac{m^2}{2E^2q^2}\left[ \omega(E+E^\prime) + \frac{m^2}{2} \right], \\
    \frac{v_0}{E^2} \hat{V}_{T^\prime} &= \frac{E_f}{E}v_{T^\prime}^{m=0} -\frac{\omega}{q} \frac{m^2}{2E^2},
\end{align}
where $m$ is the mass of the final-state lepton.
The functions $E_f/E~v_i^{m=0}$ are identical to the lepton factors in the massless case at fixed $E,E_f,q$ given in Eqs.~(\ref{eq:VCC_DE}-\ref{eq:VTp_DE}) and (\ref{eq:VCC_DE_fixE} - \ref{eq:VTp_DE_fixE}).
One sees explicitly that the lepton mass does not introduce problematic behavior or additional functional dependencies in the cross section.
They can be taken into account by including terms proportional to $\omega^n$ and/or $\mathcal{E}^n$ with $n \leq 2$.

Another thing to take into account is that the region of physical phase space shrinks. Since for fixed $q$ and $E$ the limits in the scattering angle imply $(E - \omega_{max})^2 - m_f^2 = (E \pm q)^2$. If $E$ is large compared to $q$ and  $m_f$ the shrinking of the phase space is limited $\omega_{\max} \approx q - \frac{m^2}{2(E-q)}$.

\subsection{GFMC calculations}
\label{sec:GFMC}
\begin{figure}
    \centering
    \includegraphics[width=0.48\textwidth]{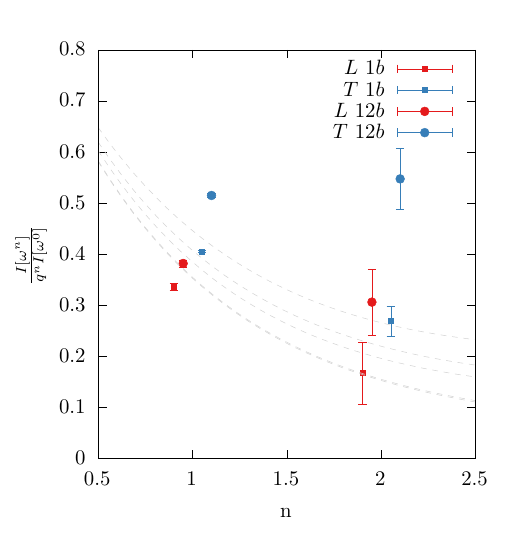}
    \caption{First and second moments computed from derivatives of the Euclidean response for electromagnetic interaction obtained in GFMC~\cite{Lovato16}. We show results for the longitudinal (red) and transverse (blue) responses, both with (circles) and without (squares) two-body currents. The points have been shifted from the central values of $n$ for ease of visibility. 
    The dashed lines show the integral of $(\omega/q)^n R(\omega)$ obtained from responses computed with the spectral function. }
    \label{fig:Moments_GFMC}
\end{figure}
In our study the high-$\omega$ tail of the nuclear response function is determined by short-range correlated pairs.
This allows one to compute corrections to Eqs.~(\ref{eq:Int_Moments}-\ref{eq:intE2}), by subtracting the strength $ q > \omega$ estimated from the nucleon momentum distribution.  
We have not considered the contribution of two-body currents or final-state interactions however, which will complicate this picture.

To illustrate this we compute the first and second moments from the Euclidean responses obtained from GFMC calculations of Ref.~\cite{Lovato16}.
These are shown in Fig.~\ref{fig:Moments_GFMC}, and are computed from third-order finite difference.
One sees that the results that include only 1-body currents scale in the same way as the PWIA results which here also include the unphysical region (that is $I^n/q^n \sim (\omega_{QE}/q)^n$). 
This is also the case for the longitudinal response when two-body currents are included, which is expected since the effect two-body currents in the electromagnetic longitudinal response is small.
When two-body currents are included the moments of the transverse response behave differently. 
The first and second moments are significantly larger (as a ratio to the 0th moment).
This is expected, two-body currents contribute strength at higher $\omega$, and the response is not solely dominated by a quasielastic peak.
It might be possible to consider responses for the deuteron, to obtain more precise information on the high-$\omega$ tail of the response, and obtain a correction suitable for the case when two-body currents are included.

On the other hand, one does not necessarily need to compute these moments from the Euclidean response.
Ab-initio calculations of Refs.~\cite{PhysRevC.89.064317,PhysRevC.94.034317, PhysRevC.105.034313} for example compute moments (and other energy-weighted sum-rules) of the nuclear responses.
And may have some control over the high-$\omega$ cut-off used in the calculation~\cite{Baccatalk}, so that strength in the unphysical region may be estimated or removed.

\subsection{Nucleon form factors}
\label{sec:ormFactors}
We have tacitly assumed that the nuclear responses and their Laplace transform include all necessary kinematic dependence on the energy transfer $\omega$. 
In GFMC calculations however the $Q^2 = q^2 - \omega^2$ dependence of the nucleon couplings, usually parametrized by form factors, is not included.
Typically the responses are computed with the couplings fixed at a central value.
When the Euclidean response is inverted, the $Q^2$ dependence of the dominant form factor can then be added to the response.

If the cross sections are computed directly from the Euclidean response, the full $\omega$-dependence of the couplings either needs to be included in the response, or we can attempt to add this dependence through the inverse Laplace transform when computing integrals of the type Eq.~(\ref{eq:laplace_fold}).
The form factors of the nucleon $G(Q^2)$ are analytic functions in the complex $Q^2$-plane with a branch cut in the left-hand plane starting at $-Q^2 = 4 m_\pi^2$.
As such, any proper parametrization of the nucleon form factor is analytic in the right-half plane, and it's inverse Laplace transform as function of $Q^2$ reduces to an inverse Fourier transform.  
However, we require the $\omega$ dependence of the form factors at fixed $q$, i.e. $G(q^2 - \omega^2)$, which now has a branch cut also in the right-hand plane.

Moreover, many parametrization of the nucleon form factors approximate the branch cut by introducing a number of poles on the real axis.
Such parametrizations are problematic to include in our procedure.
This is most easily illustrated by considering a typical dipole form
\begin{equation}
    G_D(Q^2) \propto (M_{D}^2 + q^2 - \omega^2)^{-2},
\end{equation}
where typically $M_{D} \sim 1~\mathrm{GeV}$. 
Even though the inverse Laplace transform of this form factor as function of $\omega$ is known analytically, it is problematic.
The form factor as function of $\omega$ has poles at $\omega = \sqrt{M^2 + q^2}$ and the integrals don't converge;
The inverse Laplace transform of the form factor diverges as $e^{\tau\sqrt{M_D^2 + q^2}}$ for large $\tau$.

We can avoid this since we only need the form factors in the physical region, $\omega \in [0,q]$.
As such one can approximate the $\omega$-dependence of the form factors in the physical region with a function that doesn't introduce these problematic poles.
The simplest approach is the Taylor expansion of the form factor as function of $\omega$ around the approximate position of the quasielastic peak.
The price one has to pay is that additional powers of $\omega$ are introduced, meaning that higher-order moments need to be considered.

Another option is to consider parametrizations that don't have poles on the real axis, such as the $z$-expansion of Refs.~\cite{Meyer:2016oeg,Meyer:2022mix}, which is analytic in the cut plane.
However, introducing a realistic form factor exactly will likely introduce a large amount of strength in the unphysical region $(\omega > q)$, which needs to be subtracted from the integrals.

\section{Conclusions}
\label{sec:conclusion}

Nuclear response functions are central ingredients in neutrino–nucleus cross sections relevant for accelerator-based oscillation experiments. Ab initio many-body methods can compute integral transforms of these responses with high precision, because the transform can be written as a ground-state expectation value, avoiding an explicit sum over the continuum of hadronic final states. In this work we identify situations in which, starting from the integral-transformed response, one can compute experimentally relevant cross sections without reconstructing the real-frequency response.

We show that inclusive neutrino cross sections at fixed momentum transfer $q$, when integrated over either the outgoing- or incoming-lepton energy, can be expressed as integrals over the energy transfer 
$\omega$ of the nuclear responses weighted by simple functions of 
$\omega$. This is particularly advantageous for ab initio approaches: the required quantities reduce to a small set of energy-weighted integrals of responses that can be obtained directly from the transformed response. Specifically, the cross section integrated over the incoming energy depends on moments of the response up to second order, together with integrals weighted by  $(E_f + \omega)^{-n}$ with $n \in \{0,1,2\}$. The cross section integrated over the outgoing-lepton energy is fully determined by up to the fourth moment of the responses. The integrated cross sections are linear combinations of these quantities, with kinematic coefficients given in Appendix~\ref{app:lepton_factors}. Although we consider the massless lepton limit for numerical illustrations Sec.~\ref{sec:leptonmass} shows that lepton-mass effects only modify these coefficients and do not introduce additional integrated inputs.

The energy-weighted response integrals required by our framework can be extracted directly from Euclidean responses using the relations derived in this work, provided the energy-transfer integration can be extended to sufficiently large values. As a proof of concept, we demonstrate the strategy with a Gaussian toy model for quasielastic responses, which also helps clarify the relative importance of the different integrated contributions. For the single-differential cross section at fixed momentum transfer, we show that the contributions of the moments of responses can be organized systematically in an expansion controlled by the momentum transfer relative to the nucleon mass when the response is dominated by a quasielastic peak.

This allows for calculations of this cross section that are systematically improvable by including higher order moments.
We note that the same applies of course for the equivalent cross section for electron-nucleus scattering, and that such data might be obtained in CLAS, this is explored elsewhere~\cite{Nikolakopoulos:prep_EM}.

The cross section integrated over incoming energy at 
$q = 500$ MeV, is found to be largely dominated by only two types of integrals. In particular the second order moments are strongly suppressed in this kinematic regime.
We also show that the same framework extends naturally to flux-averaged cross sections: the flux dependence can be absorbed into the kinematic weights, so the final result remains a linear combination of the same integrated response quantities, with modified coefficients. Depending on the chosen parametrization of the flux, additional higher-order moments may enter, but the overall strategy remains unchanged.

As a more realistic test, we assess the extraction of the required integrated response inputs from Euclidean data using PWIA responses based on a realistic spectral function. We include statistical uncertainties comparable to those encountered in GFMC calculations~\cite{PhysRevX.10.031068}. We find that the integrals entering our formulation can be determined with good accuracy; noticeable sensitivity to numerical noise appears mainly in higher-order moments, for which more specialized moment-evaluation techniques can be employed~\cite{PhysRevC.89.064317,PhysRevC.94.034317, PhysRevC.105.034313}.

A key issue is that the response integrals relevant for the cross section have a finite upper limit set by kinematics. When moments are inferred from Euclidean responses using formal relations that assume integration to arbitrarily large $\omega$, one must subtract contributions from the unphysical region $\omega > \omega_{\rm max}$. We argue that this correction is tractable, since it is controlled by the smooth high-$\omega$ tail of the response, which is expected to display largely universal behavior associated with short-range correlations~\cite{Tropiano:2021qgf, Ryckebusch:2019oya, Cruz-Torres:2019fum, CiofidegliAtti:2015lcu, Ryckebusch:2014ann, CLAS:2022odn}. Moreover, the unphysical tail impacts higher-order moments most strongly, but the contribution of these moments to the cross section is typically suppressed. This allows to assign conservative uncertainties on the subtraction without compromising the robustness of the final results.
We demonstrate this using a realistic spectral function where the high-$\omega$ response is dominated by the coupling to a single nucleon in a SRC pair with high relative momentum.

A further practical issue concerns single-nucleon form factors. Euclidean responses computed in GFMC do not retain the full energy dependence of these form factors, which should be accounted for in precision cross-section calculations. In principle, this dependence can be incorporated by absorbing it into the kinematic weights, but doing so exactly is nontrivial: common parametrizations (e.g., dipole forms) have an analytic structure that complicates the Euclidean-based treatment. A pragmatic option is to approximate the form factors in the physical region (for instance by Taylor expansion around the quasielastic peak), at the cost of introducing additional higher-order moments. Developing alternative representations that capture the relevant energy dependence while remaining compatible with the present framework is an interesting direction that could be investigated in a future work.

In summary, we show that energy-integrated neutrino cross sections can be obtained directly from Euclidean response information, allowing the relevant contributions to be organized systematically and the associated numerical uncertainties to be quantified in a transparent way. The central practical challenge is the subtraction of contributions from kinematically unphysical energy transfers. While we find that these corrections can be treated robustly within the approximations explored here, their stability should be reassessed once final-state interactions and two-body currents are included. Our study of electromagnetic responses indicates that one-body-current results exhibit quasielastic-dominated scaling similar to spectral-function calculations, whereas two-body currents induce noticeable deviations in the transverse response.

\section*{Acknowledgments}
We thank A. Lovato for sharing GFMC results for the Euclidean responses and discussions.
We thank S. Bacca and R. Plestid for useful discussions and comments.
This work was completed in part during the program 'Neutrino-Nucleus Interactions in the standard model and beyond' at the Mainz Institute for Theoretical Physics (MITP), whose hospitality and support are gratefully acknowledged.
AN is supported by the Neutrino Theory Network under Award Number DEAC02-07CH1135. The work of NR was funded by the European Union (ERC Starting Grant, 'NUQNET', grant agreement No. 101164195).

\appendix

\section{Lepton prefactors}
\label{app:lepton_factors}
The kinematic lepton factors, for massless (relativistic) initial and final-state leptons are~\cite{Donnelly85}
\begin{align}
v_{CC} &= 1+\cos\theta, \label{eq:VCC} \\
v_{CL} &= -2\frac{\omega}{q}v_{CC}, \label{eq:VCL} \\ 
v_{LL} &= v_{CC} - 2\frac{E E_f}{q^2} \sin^2\theta = \left(\frac{\omega}{q}\right)^2 v_{CC}, \label{eq:VLL} \\
v_T &= 2 - v_{CC} + \frac{E E_f}{q^2} \sin^2\theta, \label{eq:VT} \\
v_{T^\prime} &= \frac{E + E_l}{q}(2-v_{CC}), \label{eq:VTp}
\end{align}
in terms of the angle between incoming and outgoing lepton $\theta_f$, with $E = E_f + \omega$.
With the scattering angle written in terms of momentum transfer $q$ as
\begin{equation}
\cos\theta = \frac{q^2 - (\omega + E_f)^2 - E_f^2}{-2(\omega+E_f)E_f},
\end{equation}
one can express the relevant kinematic factors as linear combinations of powers of $\omega$ and the functions $\mathcal{E}^n(\omega) \equiv 1/(E_f+\omega)^n$:
\begin{widetext}
\begin{align}
\frac{E_f}{E} v_{CC} &= \frac{1}{2} \omega^0 + E_f \mathcal{E}^1 -\frac{q^2 - E_f^2}{2} \mathcal{E}^2, \label{eq:VCC_DE} \\
\frac{E_f}{E} v_{CL} &= - \frac{2E_f}{q} \omega^0  -\frac{1}{q} \omega^1 + \frac{q^2 + E_f^2}{q} \mathcal{E}^1 - \frac{E_f}{q} \left(q^2 - E_f^2\right)\mathcal{E}^2, \label{eq:VCL_DE} \\
\frac{E_f}{E} v_{LL} &= -\frac{q^2 + E_f^2}{2q^2} \omega^0 + \frac{E_f}{q^2} \omega^1 + \frac{1}{2q^2} \omega^2 + E_f\mathcal{E}^1 - \frac{E_f^2}{2q^2}\left(q^2 - E_f^2\right)\mathcal{E}^2, \label{eq:VLL_DE} \\
\frac{E_f}{E} v_T &= \frac{E_f^2}{4q^2} \omega^0  - \frac{E_f}{2q^2} \omega^1 - \frac{1}{4q^2} \omega^2 + E_f \mathcal{E}^1 + \frac{q^4 - E_f^4}{4q^2} \mathcal{E}^2 , \label{eq:VT_DE} \\
\frac{E_f}{E} v_{T^\prime} &= -\frac{1}{2q} \omega^1 + \frac{q^2 + E_f^2}{2q} \mathcal{E}^1 + \frac{E_f\left(q^2 - E_f^2\right)}{2q} \mathcal{E}^2. \label{eq:VTp_DE}
\end{align}
These expansion coefficients define the matrices of Eq.~(\ref{eq:defVE_D}).

When the incoming energy is fixed the lepton prefactors can be written solely in terms of powers of $\omega$ as 
\begin{align}
\frac{E_f}{E} v_{CC} &= \left( 2 - \frac{q^2}{2E^2} \right) \omega^0 - \frac{2}{E} \omega^1 + \frac{1}{2E^2}\omega^2
, \label{eq:VCC_DE_fixE} \\
\frac{E_f}{E} v_{CL} &= \left( -\frac{4}{q} + \frac{q}{E^2} \right) \omega^1 + \frac{4}{q E} \omega^2 - \frac{1}{qE^2}\omega^3 , \label{eq:VCL_DE_fixE} \\
\frac{E_f}{E} v_{LL} &= \left( \frac{2}{q^2} - \frac{1}{2E^2} \right) \omega^2 - \frac{2}{q^2 E} \omega^3 + \frac{1}{2q^2E^2}\omega^4 , \label{eq:VLL_DE_fix_E} \\
\frac{E_f}{E} v_T &= \left(1 + \frac{q^2}{4E^2} \right)\omega^0 -\frac{1}{E}\omega^1 - \frac{1}{q^2}\omega^2 + \frac{1}{Eq^2} \omega^3 - \frac{1}{4q^2 E^2} \omega^4, \label{eq:VT_DE_fix_E} \\
\frac{E_f}{E} v_{T^\prime} &= \frac{q}{E} \omega^0 - \frac{q}{2E^2} \omega^1 - \frac{1}{qE} \omega^2 + \frac{1}{2qE^2} \omega^3, \label{eq:VTp_DE_fixE}
\end{align}
which defines the matrix of Eq.~(\ref{eq:sigma_E_matrix}).
\end{widetext}
\section{Calculating moments from the Euclidean response}
\label{app:Moment_calculation}
For the calculation of the energy-integrated cross section we need to compute the $n$-th (one-sided) moment of the nuclear response
\begin{equation}
    I_i\left[ \omega^n \right] = \int_0^{\infty} \mathrm{d}\omega~\omega^{n} R_i(\omega) = E_i^{(n)}(0),
\end{equation}
corresponding to the $n$-th derivative of the Euclidean response (the moment-generating function) at $\tau=0$.
We use a simple numerical method to compute the needed lowest-order moments from the Euclidean response.
To do this we consider the following integral
\begin{equation}
\label{eq:omega_cut}
\int_0^{\infty} \mathrm{d}\omega~\frac{\omega}{1+\frac{\omega}{\Lambda}}~R_i(\omega, q) = \Lambda E_i(0)  - \Lambda^2 \int_0^{\infty} \mathrm{d}\tau E_i(\tau) e^{-\Lambda\tau},
\end{equation}
with a cut-off $\Lambda$. The right-hand side follows from Eq.~(\ref{eq:laplace_fold}) and can be directly computed from the Euclidean response $E_i(\tau)$.
For sufficiently large $\Lambda$ we can expand
\begin{align}
\label{eq:series_f1}
&\int_0^\infty \mathrm{d}~\omega \left[\omega - \frac{\omega^2}{\Lambda} + \frac{\omega^3}{\Lambda^2} + O( \Lambda^{-3} ) \right]~R_i(\omega, q)  \\
&= I_i\left[\omega^1 \right] - I_i\left[\omega^2 \right]/\Lambda + I_i\left[ \omega^3 \right]/\Lambda^2 + O(\Lambda^{-3}) 
\end{align}
as long as $R_i(\omega)$ decays sufficiently fast.
Hence the expansion coefficients that determine the $\Lambda$ dependence of Eq.~(\ref{eq:omega_cut}) are exactly the integrals we need to compute.
In addition we consider the derivative and subtract to remove the leading term
\begin{widetext}
\begin{align}
    \label{eq:omega_diff_cut}
    \int_0^{\infty} \frac{\omega^2}{(1+\frac{\omega}{\Lambda})^2} R_i(\omega)~d\omega &= \Lambda^2 \left[ E_i(0) + \Lambda \int_0^{\infty} E_i(\tau) e^{-\Lambda\tau}\left(\Lambda\tau - 2\right)\right] \\
    &\approx I_i\left[ \omega^2\right] - 2I\left[\omega^3\right]/\Lambda + 3I_i\left[\omega^4\right]/\Lambda^2 + O(\Lambda^{-3}) \label{eq:series_f2},
\end{align}
\end{widetext}
where again the first equality follows from Eq.~(\ref{eq:laplace_fold}) and the second equality from the expansion for large $\Lambda$.
We get the required integrals $I_i\left[\omega^n \right]$ by computing Eq.~(\ref{eq:omega_cut}) and Eq.~(\ref{eq:omega_diff_cut}) from the Euclidean response for different $\Lambda$. 
The $\Lambda$-dependence is then fit with inverse powers of $\Lambda$ up to $\Lambda^{-2}$, to obtain the required integrals as the expansion coefficients in Eq.~(\ref{eq:series_f1}) and (\ref{eq:series_f2}).
The fits are performed for $\Lambda \in \left[ \Lambda_{min} : \Lambda_{max}\right]$, where we fix $\Lambda_{min} = 2~\mathrm{GeV}$ and vary $\Lambda_{\max}$ to study convergence.
We compute $I_i\left[\omega^1 \right]$ from the leading fit coefficient of Eq.~(\ref{eq:series_f1}). A conservative error estimate is obtained as $I_i\left[\omega^3\right]/\Lambda_{\min}^2$ with the third moment obtained from the same fit.
The second moment $I_i\left[ \omega^2 \right]$ is obtained from the fit of Eq.~(\ref{eq:series_f2}), and again the uncertainty may estimated from the last coefficient $3I_i\left[ \omega^4 \right]/\Lambda^2_{min}$ obtained in the same fit.
These uncertainty estimates are of the order of $1\%$.
The actual discrepancy with respect to the true moments is much smaller for any $\Lambda_{max} > 3~\mathrm{GeV}$. For $\Lambda_{max}= 5~\mathrm{GeV}$ the discrepancy is at the permille level.

We used this method to compute moments from the Euclidean response in the Gaussian toy-model. As discussed in the text, the method fails when these integrals are computed on a fixed grid when the stepsize is too large $\Delta \tau\times \Lambda > 1$. It might be considered if the stepsize can be reduced or be made non-uniform, or if analytical approximations for the Euclidean response are available.  

\section{Energy dependent integrals with the factorized spectral function for $CL$, $LL$ and $T^\prime$ responses}
\label{app:Eints}
\begin{figure*}
    \includegraphics[width=0.31\textwidth]{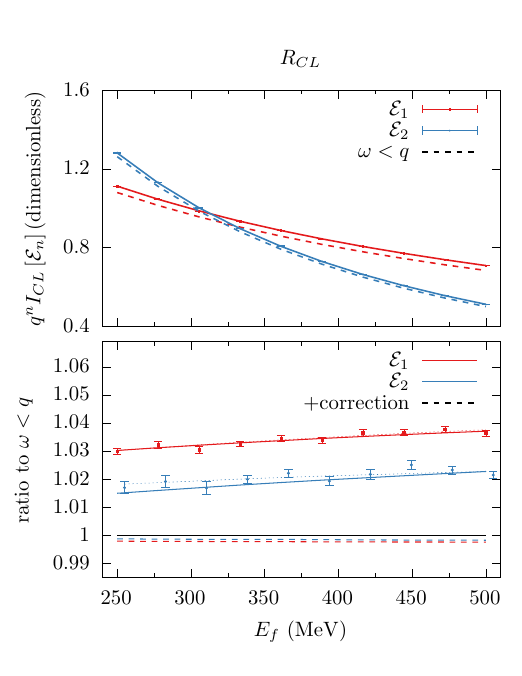}
    \includegraphics[width=0.31\textwidth]{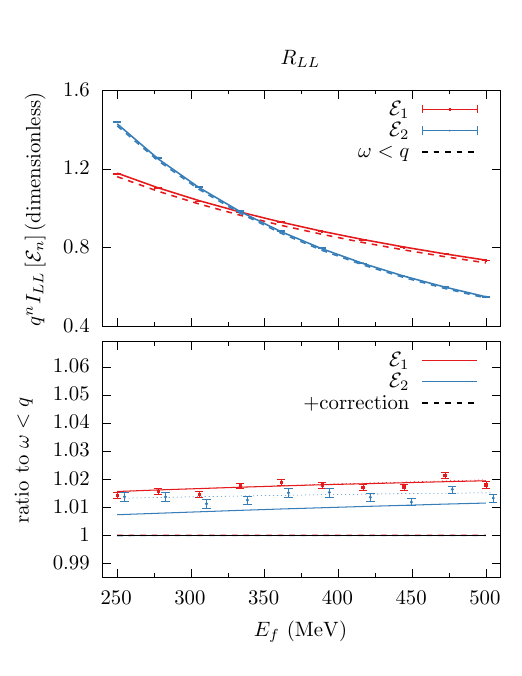}
    \includegraphics[width=0.31\textwidth]{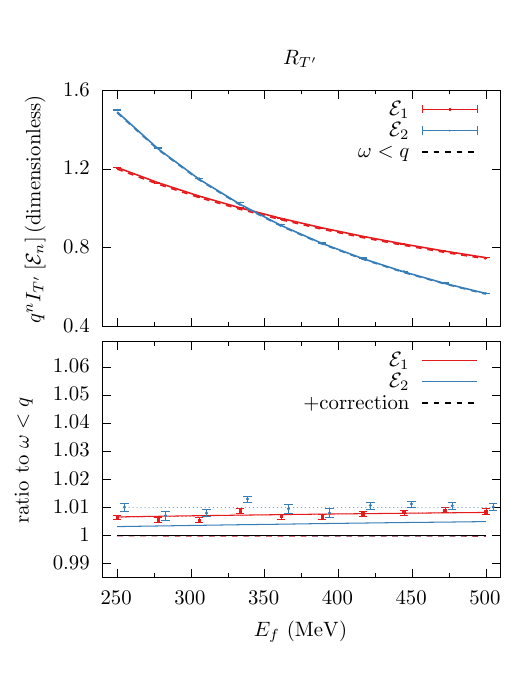}
    \caption{Same as in Fig.~\ref{fig:R_CC_E_ratio}, but for the $CL, LL,$ and $T^\prime$ responses on the left, middle and right respectively.}
    \label{fig:R_CL_LL_Tp_ratios}
\end{figure*}
The results shown for the $CC$ and $T$ responses in Figs.~\ref{fig:R_CC_E_ratio}~and~\ref{fig:R_T_E_ratio} respectively are shown in Fig.~\ref{fig:R_CL_LL_Tp_ratios} for the $CL$, $LL,$ and $T^\prime$ responses.

\section{The response at high missing momentum}
\label{app:highK_SF}
If the response at high energy is dominated by the knockout of two correlated nucleons, where the energy and momentum is transferred to a single nucleon in the pair the dominant contribution to the missing energy is given by the kinetic energy of the second nucleon. Hence the spectral function is determined by the relative momentum distribution of the nucleon pair. In this case the response can be computed from the single-nucleon momentum distribution in the PWIA.
We illustrate this here, and consider explicitly the effect of center-of-mass motion.

Following Ref.~\cite{CiofidegliAtti:1995qe}, we assume that the distribution of relative and CMS momentum factorize, and that the spectral function is given by
\begin{equation}
\label{eq:SEk_fact_CDA}
    S(E,k) = \int \mathrm{d}^3\mathbf{P}~n_{rel}(\big\lvert \mathbf{k + \frac{\mathbf{P}}{2}} \big\rvert) n_{cm}(\big\lvert \mathbf{P}\big\rvert) \delta(E-E_c -E^*_{A-1}),
\end{equation}
where the intrinsic excitation energy of the $A-1$ system is taken to be
\begin{equation}
\label{eq:T_amin1_CDA}
    E^*_{A-1} = \frac{A-2}{2M(A-1)}\left[ \mathbf{k} + \frac{A-1}{A-2}\mathbf{P} \right]^2,
\end{equation}
for $A>2$.
Note that this definition differs from $T_{A-1}$ in Eq.~(\ref{eq:SF_VMC_unfactorized}). 
Here $E^*_{A-1}$ is the kinetic energy associated with the relative motion of the single nucleon and $A-2$ system that make up the $A-1$ system.
Explicitly, with $\mathbf{P} = -(\mathbf{k} + \mathbf{k}^\prime)$
\begin{equation}
    E^*_{A-1} = \frac{\mathbf{k}^{\prime 2}}{2M} + \frac{\left(\mathbf{k}^\prime + \mathbf{k}\right)^2}{2M(A-2)} - \frac{\mathbf{k}^2}{2M(A-1)},
\end{equation}
so that $E_{A-1}^* = T_{A-1} - \frac{\mathbf{k}^2}{2M(A-1)}$, i.e. the difference is the recoil energy of a $A-1$ system.
In the following we use this definition from Ref.~\cite{CiofidegliAtti:1995qe}.
One can add the recoil term in the missing energy in the results obtained from Eq.~(\ref{eq:T_amin1_CDA}) to obtain the kinetic energy of the residual system in the LAB frame $T_{A-1}$.

We consider this model first in the case where $n_{cm}$ is a delta function, then we consider a semi-analytical example where $n_{cm}$ and $n_{rel}$ are both Gaussian.
This toy model case suggests corrections which we compare to numerical results with momentum distributions obtained in VMC calculations.

\subsection{$n_{cm}(\mathbf{P}) = \delta(\mathbf{P})$}
In this case\footnote{If one uses $T_{A-1}$ instead of $E^*_{A-1}$ as discussed before, $(A-2)/(A-1) \rightarrow 1$.}
\begin{equation}
    S(E,k) = n_{rel}(k)~\delta\left( T - \frac{A-2}{A-1} \frac{k^2}{2M}\right),
\end{equation}
where $T\equiv E - E_c$.
The energy and momentum distributions are
\begin{equation}
\label{eq:nk_deltaf}
    n(k) = \int \mathrm{d} E~S(E,k) = n_{rel}(k),  
\end{equation}
and
\begin{align}
\label{eq:SE_deltaf}
    S(E) &= 4\pi \int \mathrm{d}k~k^2~S(E,k) \\ 
    &= 4\pi M \left(\frac{A-1}{A-2}\right) k_0(E)~n_{rel}\left( k_0(E)\right) \theta(T),
\end{align}
where we use the shorthand
\begin{equation}
    k_0(E) = \sqrt{2M(E-E_c)\frac{A-1}{A-2}}.
\end{equation}
The response is 
\begin{align}
    &R_{PS}(\omega,q) = 2\pi \frac{M}{q} \int_{0}^{\omega} \mathrm{d} E \int_{k^-(E,\omega,q)}^{k^+(E,\omega,q)} \mathrm{d} k~kS(E,k) \nonumber \\ 
    &= 2\pi \frac{M^2}{q}\frac{A-1}{A-2}\int_{T^-(\omega,q)}^{T+(\omega,q)}\mathrm{d}T~ n_{rel}(\sqrt{2MT\frac{A-1}{A-2}}).
    \label{eq:R_PS_delta}
\end{align}
The bounds on the energy integral are given by the requirement that 
\begin{equation}
    \label{eq:ineq_bounds}
    k^-(\omega,q,T) < \sqrt{2MT\frac{A-1}{A-2}} < k^+(\omega,q,T).
\end{equation}
If one treats the outgoing nucleon non-relativistically, the bounds on the missing momentum integral are given by
\begin{equation}
\label{k_minmax_nr}
    k^{\pm}(\omega,q,T) = \big\lvert \sqrt{2M(\tilde{\omega} - T)} \pm q \big\rvert.
\end{equation}
were we introduced $\tilde{\omega} = \omega - E_c$.
The non-relativistic treatment is suitable for typical ab-initio calculations.
If the outgoing nucleon is treated in relativistic kinematics, as we do in this work, $k^{\pm}(\omega,q,T) = \lvert \sqrt{ (\tilde{\omega} - T) \left( \tilde{\omega} - T + 2M\right)} \pm q\rvert$.
The explicit expression for the bounds $T^{\pm}(\omega,q)$ is somewhat unwieldy in this case, but easily found numerically.

The integral of Eq.~(\ref{eq:R_PS_delta}) is in more familiar form
\begin{equation}
\label{eq:R_PS_scaling}
        R_{PS}(\omega,q)= 2\pi \frac{M}{q} \int_{y^{-}(\omega,q)}^{y^{+}(\omega,q)} k\mathrm{d}k~n(k),
\end{equation}
where $y^{\pm}(\omega,q) \equiv \sqrt{2MT^{\pm}(\omega,q) \frac{A-1}{A-2}}$ and $n(k)$ is the (relative) momentum distribution.
In typical scaling analysis one takes the large $q$-limit, and approximates the lower-bound. 

For the purpose of the next section, where center-of-mass motion is taken into account, it is useful to introduce an arbitrary scaling of the inequality of Eq.~(\ref{eq:ineq_bounds}).
That is we define $y^{\pm}(\omega,q,s)$ as the bounds on $k_0$ determined from the inequality
\begin{equation}
\label{eq:ineq_scaled}
k^-(\omega,q,k_0) < sk_0 < k^+(\omega,q,k_0),
\end{equation}
i.e.,
\begin{equation}
    k^{\pm}(\omega,q,y^{\pm}) = sy^\pm(\omega,q,s).
\end{equation}
So that $y^{\pm}(\omega,q,s)$ are the solutions to depressed quartics.

\subsection{Gaussians}
\begin{figure}
    \centering
    \includegraphics[width=0.48\linewidth]{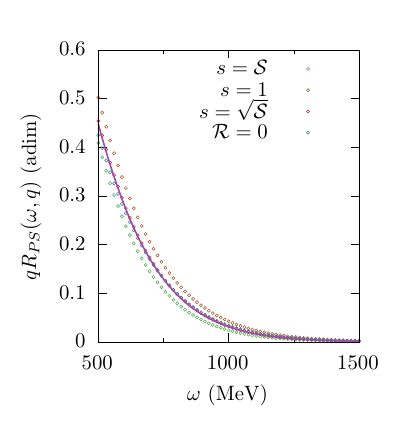}
    \includegraphics[width=0.48\linewidth]{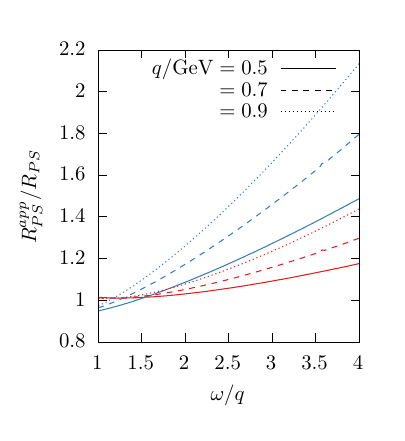}
    \caption{(left:) Comparison of the response from the two-Gaussian spectral function at $q=500$ to approximations of Eq.~\ref{eq:2Gauss_napprox}. The results labeled with $s= \mathcal{S},~1,~\sqrt{\mathcal{S}}$ correspond to modifications of the integration bounds as in Eq.~(\ref{eq:ineq_scaled}). Here $\mathcal{S} = \frac{1-\frac{A}{A-2}\mathcal{R}}{1+\mathcal{R}}$ based on Eq.~(\ref{eq:kcm_0_errorfunc}). The result labeled $\mathcal{R}=0$ is the 'typical approximation', taking $\mathcal{R}=0$ in Eq.~(\ref{eq:2Gauss_napprox}). 
    (right:) Ratio of the approximations to the exact result for different values of $q$. Here red lines correspond to the $s=\sqrt{\mathcal{S}}$ approximation while blue lines to the $\mathcal{R}=0$ approximation.}
    \label{fig:R_comp_Gauss}
\end{figure}

We will assume that both the relative and center-of-mass momentum distributions are Gaussian, which allows for partly analytic results.
Such a model has for example been used in Refs.~\cite{JeffersonLabHallA:2022ljj, JeffersonLabHallA:2022cit}.
We define
\begin{equation}
    n_{rel}(p)~n_{cm}(P) = N~\left(\pi k_{r} k_{cm}\right)^{-3} e^{\left[ -\left(\frac{p}{k_{r}}\right)^2 - \left(\frac{P}{k_{cm}}\right)^2 \right]},
\end{equation}
where $N$ is the overall normalization of the spectral function.
The spectral function can then be written as
\begin{equation}
S(E,k) = \mathcal{N} \frac{M}{k}~e^{-a k^2} e^{-b k_0^2} \sinh\left(c~k_0 k\right) \theta(T)
\end{equation}
where again $T = E - E_c$, and $k_0 = \sqrt{2MT\frac{A-1}{A-2}}$~\footnote{If one uses $T_{A-1}$ instead of $E^*_{A-1}$ one replaces $T \rightarrow T+ \frac{k^2}{2M(A-1)}$. One has $k_0 = \sqrt{2MT\frac{A-1}{A-2}}\left[ 1 + \frac{k^2}{2MT(A-1) } - \ldots \right]$. Since the spectral function peaks around $T\sim k^2/(2M)$ these corrections are small for any $k$.}. Here
\begin{equation}
a = \left(\frac{A-2}{A-1}\right)^2\left[ \frac{1}{k_{cm}^2} + \left(\frac{ A}{A-2}\right)^2\frac{1}{4 k_{r}^2} \right],
\end{equation}
\begin{equation}
b = \left(\frac{A-2}{A-1}\right)^2 \left[ \frac{1}{k_{cm}^2} + \frac{1}{4k_{r}^2} \right],
\end{equation}
\begin{equation}
c = 2\left( \frac{A-2}{A-1}\right)^{2} \left[\frac{1}{k_{cm}^2} - \frac{A}{4(A-2)} \frac{1}{k_{r}^2} \right],
\end{equation}
\begin{equation}
    \mathcal{N} = \frac{16}{\pi^2} \frac{1}{2k_r k_{cm}\left[ (2k_r)^2 - \frac{A}{A-2}k_{cm}^2 \right]}.
\end{equation}
The momentum and energy distributions can be written in terms of the relative momentum distribution as
\begin{equation}
\label{eq:ndist_Gauss}
n(k) = \frac{ N }{\left[ 1 + \left(\frac{A-2}{A}\right)^2\mathcal{R}\right]^{3/2} }~n_{rel}\left( \frac{k} {\sqrt{1+ \left(\frac{A-2}{A}\right)^{2}\mathcal{R}}} \right),
\end{equation}
and
\begin{equation}
    S(E) = N~ \frac{4\pi M \frac{A-1}{A-2} k_0}{\left[ 1 + \mathcal{R}\right]^{3/2}} n_{rel} \left( \frac{k_0}{\sqrt{1 + \mathcal{R}}} \right) \theta(T).
\end{equation}
Where the ratio
\begin{equation}
    \mathcal{R} \equiv \left(\frac{A}{A-2} \frac{k_{cm}}{2k_{r}}\right)^2,
\end{equation}
is introduced.
Evidently, when $k_{cm} = 0$, one recovers Eqs.~(\ref{eq:nk_deltaf}-\ref{eq:SE_deltaf}).
The deviation from the delta function result are determined by $k^2_{cm}/(2k_{rel})^2$.  
Typically $2k_{cm}/3 \approx k_F$ is of the order of the Fermi momentum and $2k_{r}/3 \approx 600~\mathrm{MeV}$ is the large relative momentum scale~\cite{CiofidegliAtti:1995qe, JeffersonLabHallA:2022cit}. 
Hence $\mathcal{R}$ yields percent level deviations from the delta function result. The fact that the momentum scale $k_{cm}$ enters with a factor $1/2$ is due to Eq.~(\ref{eq:SEk_fact_CDA}).

\begin{widetext}
The response is
\begin{equation}
    R_{PS}(\omega,q) = \frac{M}{q}  \int_0^{\omega-E_c} \mathrm{d} T~ \frac{1+\mathcal{R}}{1-\frac{A-2}{A}\mathcal{R}}\frac{S(T+E_c)}{4 k_0}  \left[ \mathrm{erf}\left( \sqrt{a}  x - k_0\frac{c}{2\sqrt{a}} \right) - \mathrm{erf}\left(\sqrt{a}  x + k_0\frac{c}{2\sqrt{a}} \right) \right]_{k^{-}(\omega,q,k_0)}^{k^+(\omega,q,k_0)}.
\end{equation}
The error functions act like softened step-functions enforcing the bounds $k^- \lesssim \frac{c}{2a}k_0 \lesssim k^+$.
Indeed, the $k_{cm} = 0$ limit is
\begin{align}
    \lim_{k_{cm} \rightarrow 0} \mathrm{erf}\left( \sqrt{a} k^{\pm} \pm k_0\frac{c}{2\sqrt{a}}\right) = \lim_{k_{cm}\rightarrow 0} \mathrm{erf}\left( \frac{A-2}{A-1} \frac{\sqrt{1+\mathcal{R}}}{k_{cm}}\left[ k^{\pm} \pm \frac{1-\frac{A}{A-2}\mathcal{R}}{1+\mathcal{R} } k_0 \right] \right) 
    = 2\theta\left(k^\pm  \pm k_0\right) - 1
    \label{eq:kcm_0_errorfunc}
\end{align}
\end{widetext}
By comparison to the $k_{cm}=0$ limit, one sees two competing effects which partly cancel out.
The softening of the bounds will move some strength to smaller energies, thereby increasing the response.
On the other hand, $k_0$ is scaled with a factor $\mathcal{S} \equiv c/(2a)=\frac{1-\frac{A}{A-2}\mathcal{R}}{1+\mathcal{R} } < 1$, which moves the lower bound to larger $k_0$, thereby decreasing the response.

If we neglect the softening of the bounds to approximate the response as an integral over the momentum distribution we have
\begin{align}
R_{PS}^{app}(\omega,q) &= 2\pi \frac{M}{q} \frac{\sqrt{(1+\mathcal{R})(1 + (\frac{A-2}{A})^2\mathcal{R})}}{1-\frac{A-2}{A}\mathcal{R}}  \nonumber \\
&\times\int_{\mathcal{Y}^{-}(\omega,q,s)}^{\mathcal{Y}^+(\omega,q,s)} k\mathrm{d}k~n(k),
\label{eq:2Gauss_napprox}
\end{align}
where $n(k)$ is the momentum distribution, and the bounds are given by
\begin{equation}
    \mathcal{Y}^{\pm}(\omega,q,s) = y^{\pm}(\omega,q,s) \sqrt{\frac{1+(\frac{A-2}{A})^2\mathcal{R}}{1+\mathcal{R}}}.
\end{equation}
I.e. we approximated the error functions by step functions, but we retained the possibility of a shift $s$ of the bounds as in Eq.~(\ref{eq:ineq_scaled}).

The effect of shifting the bounds is shown explicitly in Fig.~\ref{fig:R_comp_Gauss}.
As expected, taking $s=\mathcal{S}$ as suggested by Eq.~(\ref{eq:kcm_0_errorfunc}), underpredicts the response.
Taking the $k_{cm}=0$ limit for this scaling factor $s=1$, overpredicts the response.
The softening of the bounds compensates for this.
The `typical approach', i.e. taking $\mathcal{R}=0$ (implying of course $s=1$) in Eq.~(\ref{eq:2Gauss_napprox}) but keeping $n(k)$ as the full single-nucleon momentum distribution of Eq.~(\ref{eq:ndist_Gauss}) is found to perform much better.
We find, but have not derived, that `splitting the difference' by taking $s = \sqrt{\mathcal{S}}$ provides an even better approximation to the response.

This is shown in the right panel of Fig.~\ref{fig:R_comp_Gauss}, where the ratio to the exact result is shown for different values of $q$.
Evidently, the leading order deviation for the 'typical approach' is linear in $\omega/q$ with slope dependent on q.
On the other hand, the deviation for the result with $s=\sqrt{S}$ seems to be to first order quadratic in $\omega/q$
\footnote{Note that the ratio to the full result in the $\mathcal{R}=0$ case depends (approximately linearly) on $q$. One can not obtain such a correction by scaling the integration bounds obtained with $s=1$ by a factor independent of $\omega,q$.
On the other hand, introducing a constant parameter $s$ introduces $q,\omega$-dependent shifts of the integration bounds.}.
It seems possible that a different choice for $s$ might reduce the deviation even further.

\subsection{Numerical examples with VMC momentum distributions}
\begin{figure}
    \centering
    \includegraphics[width=0.22\textwidth]{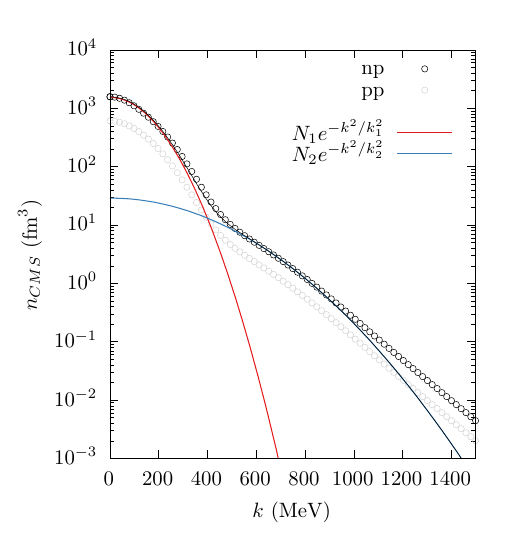}
    \includegraphics[width=0.22\textwidth]{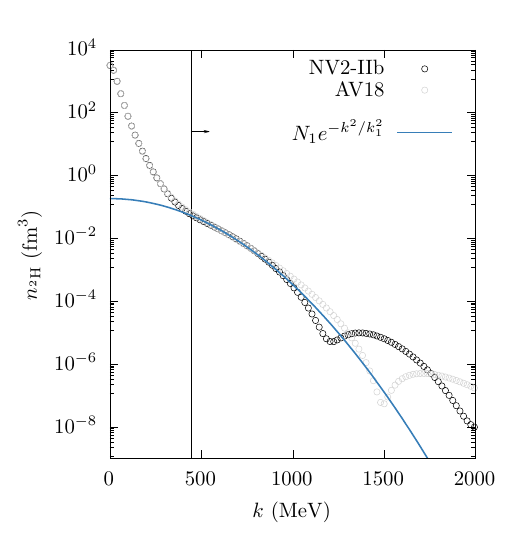}
    \caption{CMS (left) and relative (right) momentum distributions used in the numerical examples along with Gaussian fits as explained in the text.}
    \label{fig:distributions_VMC}
\end{figure}
We now consider numerical results using the factorized form of Eq.~(\ref{eq:SEk_fact_CDA}).
We use VMC results obtained for $^{12}$C for $n_{cm}(\mathbf{P})$, and approximate $n_{rel}(\mathbf{k})$ by the deuteron momentum distribution obtained with the same AV18 force~\cite{PhysRevC.89.024305, nkk_web, PhysRevC.51.38}.
The distributions are shown in Fig.~\ref{fig:distributions_VMC}, including also the deuteron momentum distribution obtained in Ref.~\cite{PhysRevC.107.014314}.
Since the results here are per illustration, we use only the $np$ CMS distribution, the shape of the $pp$ distribution is similar.

We can make a simple estimate of the effect of the relative momentum distribution based on the results in the previous section.
As is known, the low-$\mathbf{P}$ part of $n_{cm}$ is well-approximated by a Gaussian~\cite{CiofidegliAtti:1995qe}.
The figure shows a two-Gaussian approximation 
\begin{equation}
    n_{cm}(k) \approx N_{cm} (\pi k_{cm})^{-3/2} e^{-\frac{k^2}{k_{cm}^2}} + N_2 (\pi k_2)^{-3/2} e^{-\frac{k^2}{k_2^2}}.
\end{equation}
The result shown corresponds to $k_{cm} = 183~\mathrm{MeV}$, $k_2 = 450~\mathrm{MeV}$.

To determine a relevant estimate of $k_r$, the momentum scale for the relative momentum, we consider large-$\omega$, so that we are only sensitive to the high-momentum tail of $n_{rel}(k)$.
E.g. for $q=500$ MeV, at the onset of the unphysical region $\omega = q$, the lower bound of the integral in Eq.~(\ref{eq:R_PS_scaling}) is approximately $y^-(\omega,q) \approx 440~\mathrm{MeV}$.
This bound is indicated in Fig.~\ref{fig:distributions_VMC} , and a Gaussian fit to the momentum distribution above this value is shown.
One finds $k_r = 398$ MeV.
Hence the ratio $(k_{cm}/(2k_r))^2 \approx 0.05$. 
For the contribution of the tail, we find $N_2/N_{cm} \approx 0.08$. 

\begin{figure*}
    \includegraphics[width=0.3\textwidth]{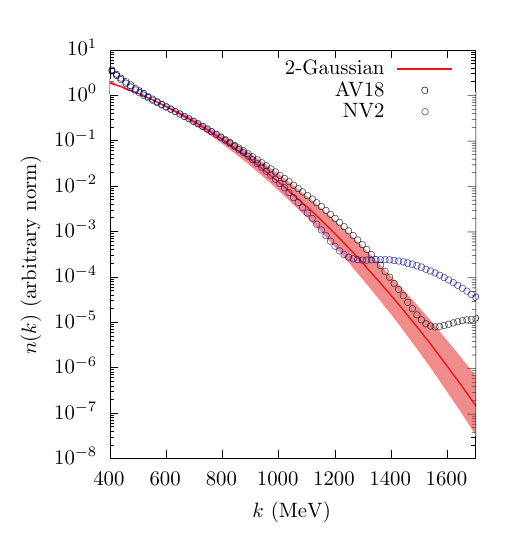} 
    \includegraphics[width=0.3\textwidth]{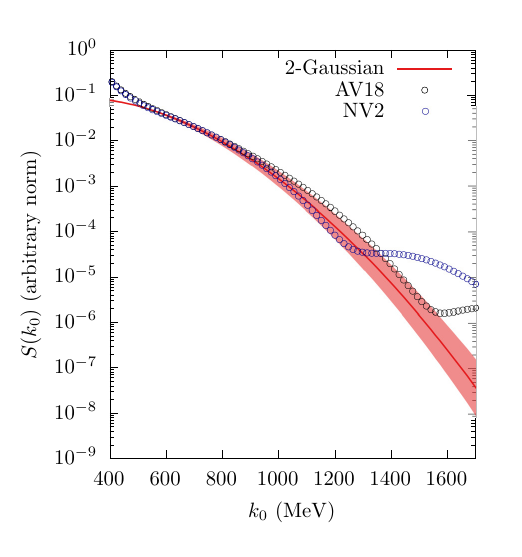} 
    \includegraphics[width=0.3\textwidth]{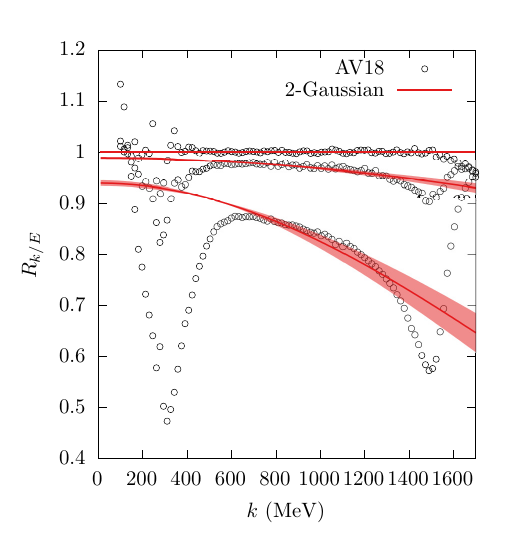} 
    \caption{Momentum and energy distributions obtained from the VMC momentum distributions compared to those from the two-Gaussian model. The two Gaussian model results are multiplied by a factor $1.2$. The energy distribution is presented as function of $k_0$, defined as $k_0^2 = 2M\frac{A-1}{A-2}(E+E_c)$ where $E$ is the missing energy. The rightmost panel shows the ratio defined in Eq.~(\ref{eq:Ratioke}) for values $A=12, 60, \infty$ (from bottom to top).}
    \label{fig:E_N_VMC_2G}
\end{figure*}
In Fig.~\ref{fig:E_N_VMC_2G} we show the momentum and energy distributions obtained with the factorized form using VMC distributions.
These are compared (shape only) to the two-Gaussian approximation using the parameters above, where the high-momentum tail of $n_{cm}$ is neglected (i.e. $N_2 = 0$).
This simple model is seen to provide a reasonable description of the momentum distribution in the region of high $k,E$. Hence one expects the approximations for the response obtained in the 2-Gaussian model to be reasonable.
The error bar for the $2$-Gaussian model is obtained by varying the parameter $k_{r}$ by 10 percent, while keeping the overall normalization the same.
Here the energy distribution is presented as function of $k_0$ defined above. I.e. we present $S(E(k))$ where $E(k) \equiv \frac{k^2}{2M}\frac{A-2}{A-1} + E_c$ with $E_c=25~\mathrm{MeV}$ the fixed threshold energy.

One of the results from the last sections is that the ratio   
\begin{equation}
\label{eq:Ratioke}
     R_{k/E}(k) \equiv  \frac{4\pi M k~n(k)}{S\left(E(k)\right)},
\end{equation} 
is constant  when $k_{cm}\rightarrow 0$ (it is $(A-1)/(A-2)$), and is unity in the large-$A$ limit independent of $k$
\begin{equation}
    \lim_{A\rightarrow\infty} R_{k/E} = 1.
\end{equation}
This latter limit can be directly derived from Eq.~(\ref{eq:SEk_fact_CDA}), and is independent of the functional forms of relative and cm momentum distributions.
The $k$-dependence of $R_{k/E}$ is instead determined by the energy-momentum relation $E^*_{A-1}(\mathbf{k}, \mathbf{P})$. 
This can be generalized to finite $A$ and to arbitrary functions for $E^*_{A-1}$, which will be explored elsewhere.
The ratio is shown in the rightmost panel of Fig.~\ref{fig:E_N_VMC_2G} for different $A$ using the VMC momentum distributions (keeping the function $n_{cm}$ the same in each case).
For $A=12$ large deviations from the Gaussian model occur at low-$k$ and around the position of the dip in the momentum distribution.

\begin{figure}
    \centering
    \includegraphics[width=0.49\linewidth]{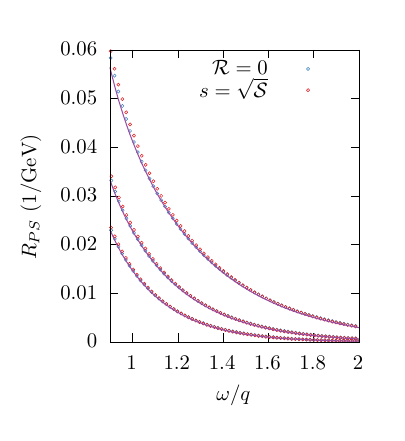}
    \includegraphics[width=0.49\linewidth]{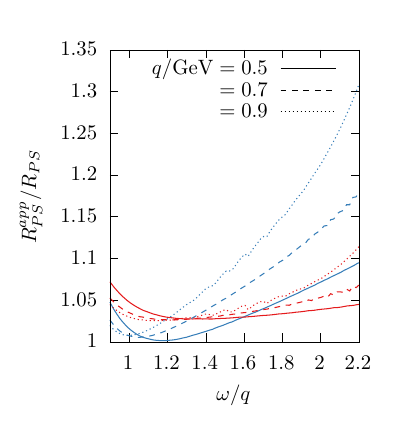}
    \caption{Comparison of the response obtained from the factorized spectral function at $q=500,~700,~900~\mathrm{MeV}$ (solid lines) to the approximations based on integrals over the momentum distribution (circles). The right panel shows the ratio of the approximations to the full result for the $\mathcal{R}=0$ model in blue and the $s=\sqrt{S}$ model in red.}
    \label{fig:Resp_comp_VMC}
\end{figure}
Finally, responses $R_{PS}$ obtained at different q are shown in Fig~\ref{fig:Resp_comp_VMC}.
The results are compared to the approximation of Eq.~(\ref{eq:2Gauss_napprox}) where we use the single nucleon momentum distribution $n(k)$ obtained from the VMC spectral function.
We compare the full result to the $\mathcal{R}=0$ approximation, i.e. Eq.~(\ref{eq:R_app_ndist}), and the $s=\sqrt{S}$ result derived in the two-Gaussian case, Eq.~(\ref{eq:2Gauss_napprox}).
As discussed in the previous section, the latter takes into account a shift in the missing energy due to the cm motion.
We use a fixed value of $\mathcal{R}$, obtained from the Gaussian fits of $k_{cm}$ and $k_r$.

From the direct comparison of the responses, Fig.~\ref{fig:Resp_comp_VMC}, it seems that the $\mathcal{R}=0$ approximation performs better in this case.
However, when one looks at the ratios with respect to the full result, a similar picture to the two Gaussian case shown in Fig.~\ref{fig:R_comp_Gauss} emerges.
The approximation that neglects center of mass motion deviates from the exact result linearly in $\omega$ with a slope depending on $q$.
For the $s=\sqrt{S}$ approximation the $q$-dependence of the deviation is again much reduced, i.e. the shape of the response is reproduced more accurately.
The normalization is off by approximately $4\%$. 
This is quite reasonable, given that the parameter $\mathcal{R}$ that gives corrections to the normalization was simply determined from a Gaussian fit.
Varying the normalization and the shift $s$ independently would improve agreement.

\bibliographystyle{apsrev4-1.bst}
\bibliography{bibliography}

\begin{thebibliography}{51}%
\makeatletter
\providecommand \@ifxundefined [1]{%
 \@ifx{#1\undefined}
}%
\providecommand \@ifnum [1]{%
 \ifnum #1\expandafter \@firstoftwo
 \else \expandafter \@secondoftwo
 \fi
}%
\providecommand \@ifx [1]{%
 \ifx #1\expandafter \@firstoftwo
 \else \expandafter \@secondoftwo
 \fi
}%
\providecommand \natexlab [1]{#1}%
\providecommand \enquote  [1]{``#1''}%
\providecommand \bibnamefont  [1]{#1}%
\providecommand \bibfnamefont [1]{#1}%
\providecommand \citenamefont [1]{#1}%
\providecommand \href@noop [0]{\@secondoftwo}%
\providecommand \href [0]{\begingroup \@sanitize@url \@href}%
\providecommand \@href[1]{\@@startlink{#1}\@@href}%
\providecommand \@@href[1]{\endgroup#1\@@endlink}%
\providecommand \@sanitize@url [0]{\catcode `\\12\catcode `\$12\catcode
  `\&12\catcode `\#12\catcode `\^12\catcode `\_12\catcode `\%12\relax}%
\providecommand \@@startlink[1]{}%
\providecommand \@@endlink[0]{}%
\providecommand \url  [0]{\begingroup\@sanitize@url \@url }%
\providecommand \@url [1]{\endgroup\@href {#1}{\urlprefix }}%
\providecommand \urlprefix  [0]{URL }%
\providecommand \Eprint [0]{\href }%
\providecommand \doibase [0]{http://dx.doi.org/}%
\providecommand \selectlanguage [0]{\@gobble}%
\providecommand \bibinfo  [0]{\@secondoftwo}%
\providecommand \bibfield  [0]{\@secondoftwo}%
\providecommand \translation [1]{[#1]}%
\providecommand \BibitemOpen [0]{}%
\providecommand \bibitemStop [0]{}%
\providecommand \bibitemNoStop [0]{.\EOS\space}%
\providecommand \EOS [0]{\spacefactor3000\relax}%
\providecommand \BibitemShut  [1]{\csname bibitem#1\endcsname}%
\let\auto@bib@innerbib\@empty
\bibitem [{\citenamefont {Alvarez-Ruso}\ \emph {et~al.}(2018)\citenamefont
  {Alvarez-Ruso}, \citenamefont {Sajjad Athar}, \citenamefont {Barbaro},
  \citenamefont {Cherdack}, \citenamefont {Christy}, \citenamefont {Coloma},
  \citenamefont {Donnelly}, \citenamefont {Dytman}, \citenamefont
  {de Gouv\^{e}a}, \citenamefont {Hill}, \citenamefont {Huber}, \citenamefont
  {Jachowicz}, \citenamefont {Katori}, \citenamefont {Kronfeld}, \citenamefont
  {Mahn}, \citenamefont {Martini}, \citenamefont {Morf\'{i}n}, \citenamefont
  {Nieves}, \citenamefont {Perdue}, \citenamefont {Petti}, \citenamefont
  {Richards}, \citenamefont {S\'{a}nchez}, \citenamefont {Sato}, \citenamefont
  {Sobczyk},\ and\ \citenamefont {Zeller}}]{NUSTECWP}%
  \BibitemOpen
  \bibfield  {author} {\bibinfo {author} {\bibfnamefont {L.}~\bibnamefont
  {Alvarez-Ruso}}, \bibinfo {author} {\bibfnamefont {M.}~\bibnamefont
  {Sajjad Athar}}, \bibinfo {author} {\bibfnamefont {M.}~\bibnamefont
  {Barbaro}}, \bibinfo {author} {\bibfnamefont {D.}~\bibnamefont {Cherdack}},
  \bibinfo {author} {\bibfnamefont {M.}~\bibnamefont {Christy}}, \bibinfo
  {author} {\bibfnamefont {P.}~\bibnamefont {Coloma}}, \bibinfo {author}
  {\bibfnamefont {T.}~\bibnamefont {Donnelly}}, \bibinfo {author}
  {\bibfnamefont {S.}~\bibnamefont {Dytman}}, \bibinfo {author} {\bibfnamefont
  {A.}~\bibnamefont {de Gouv\^{e}a}}, \bibinfo {author} {\bibfnamefont
  {R.}~\bibnamefont {Hill}}, \bibinfo {author} {\bibfnamefont {P.}~\bibnamefont
  {Huber}}, \bibinfo {author} {\bibfnamefont {N.}~\bibnamefont {Jachowicz}},
  \bibinfo {author} {\bibfnamefont {T.}~\bibnamefont {Katori}}, \bibinfo
  {author} {\bibfnamefont {A.}~\bibnamefont {Kronfeld}}, \bibinfo {author}
  {\bibfnamefont {K.}~\bibnamefont {Mahn}}, \bibinfo {author} {\bibfnamefont
  {M.}~\bibnamefont {Martini}}, \bibinfo {author} {\bibfnamefont
  {J.}~\bibnamefont {Morf\'{i}n}}, \bibinfo {author} {\bibfnamefont
  {J.}~\bibnamefont {Nieves}}, \bibinfo {author} {\bibfnamefont
  {G.}~\bibnamefont {Perdue}}, \bibinfo {author} {\bibfnamefont
  {R.}~\bibnamefont {Petti}}, \bibinfo {author} {\bibfnamefont
  {D.}~\bibnamefont {Richards}}, \bibinfo {author} {\bibfnamefont
  {F.}~\bibnamefont {S\'{a}nchez}}, \bibinfo {author} {\bibfnamefont
  {T.}~\bibnamefont {Sato}}, \bibinfo {author} {\bibfnamefont {J.}~\bibnamefont
  {Sobczyk}}, \ and\ \bibinfo {author} {\bibfnamefont {G.}~\bibnamefont
  {Zeller}},\ }\href
  {http://www.sciencedirect.com/science/article/pii/S0146641018300061}
  {\bibfield  {journal} {\bibinfo  {journal} {Progress in Particle and Nuclear
  Physics}\ }\textbf {\bibinfo {volume} {100}},\ \bibinfo {pages} {1} (\bibinfo
  {year} {2018})}\BibitemShut {NoStop}%
\bibitem [{\citenamefont {Ruso}\ \emph {et~al.}(2022)\citenamefont {Ruso} \emph
  {et~al.}}]{Ruso:2022qes}%
  \BibitemOpen
  \bibfield  {author} {\bibinfo {author} {\bibfnamefont {L.~A.}\ \bibnamefont
  {Ruso}} \emph {et~al.},\ }\href@noop {} {\  (\bibinfo {year} {2022})},\
  \Eprint {http://arxiv.org/abs/2203.09030} {arXiv:2203.09030 [hep-ph]}
  \BibitemShut {NoStop}%
\bibitem [{\citenamefont {Carlson}\ \emph {et~al.}(2015)\citenamefont
  {Carlson}, \citenamefont {Gandolfi}, \citenamefont {Pederiva}, \citenamefont
  {Pieper}, \citenamefont {Schiavilla}, \citenamefont {Schmidt},\ and\
  \citenamefont {Wiringa}}]{RevModPhys.87.1067}%
  \BibitemOpen
  \bibfield  {author} {\bibinfo {author} {\bibfnamefont {J.}~\bibnamefont
  {Carlson}}, \bibinfo {author} {\bibfnamefont {S.}~\bibnamefont {Gandolfi}},
  \bibinfo {author} {\bibfnamefont {F.}~\bibnamefont {Pederiva}}, \bibinfo
  {author} {\bibfnamefont {S.~C.}\ \bibnamefont {Pieper}}, \bibinfo {author}
  {\bibfnamefont {R.}~\bibnamefont {Schiavilla}}, \bibinfo {author}
  {\bibfnamefont {K.~E.}\ \bibnamefont {Schmidt}}, \ and\ \bibinfo {author}
  {\bibfnamefont {R.~B.}\ \bibnamefont {Wiringa}},\ }\href {\doibase
  10.1103/RevModPhys.87.1067} {\bibfield  {journal} {\bibinfo  {journal} {Rev.
  Mod. Phys.}\ }\textbf {\bibinfo {volume} {87}},\ \bibinfo {pages} {1067}
  (\bibinfo {year} {2015})}\BibitemShut {NoStop}%
\bibitem [{\citenamefont {Leidemann}\ and\ \citenamefont
  {Orlandini}(2013)}]{LEIDEMANN2013158}%
  \BibitemOpen
  \bibfield  {author} {\bibinfo {author} {\bibfnamefont {W.}~\bibnamefont
  {Leidemann}}\ and\ \bibinfo {author} {\bibfnamefont {G.}~\bibnamefont
  {Orlandini}},\ }\href {\doibase https://doi.org/10.1016/j.ppnp.2012.09.001}
  {\bibfield  {journal} {\bibinfo  {journal} {Progress in Particle and Nuclear
  Physics}\ }\textbf {\bibinfo {volume} {68}},\ \bibinfo {pages} {158}
  (\bibinfo {year} {2013})}\BibitemShut {NoStop}%
\bibitem [{\citenamefont {Lovato}\ \emph {et~al.}(2016)\citenamefont {Lovato},
  \citenamefont {Gandolfi}, \citenamefont {Carlson}, \citenamefont {Pieper},\
  and\ \citenamefont {Schiavilla}}]{Lovato16}%
  \BibitemOpen
  \bibfield  {author} {\bibinfo {author} {\bibfnamefont {A.}~\bibnamefont
  {Lovato}}, \bibinfo {author} {\bibfnamefont {S.}~\bibnamefont {Gandolfi}},
  \bibinfo {author} {\bibfnamefont {J.}~\bibnamefont {Carlson}}, \bibinfo
  {author} {\bibfnamefont {S.~C.}\ \bibnamefont {Pieper}}, \ and\ \bibinfo
  {author} {\bibfnamefont {R.}~\bibnamefont {Schiavilla}},\ }\href {\doibase
  10.1103/PhysRevLett.117.082501} {\bibfield  {journal} {\bibinfo  {journal}
  {Phys. Rev. Lett.}\ }\textbf {\bibinfo {volume} {117}},\ \bibinfo {pages}
  {082501} (\bibinfo {year} {2016})}\BibitemShut {NoStop}%
\bibitem [{\citenamefont {Lovato}\ \emph {et~al.}(2020)\citenamefont {Lovato},
  \citenamefont {Carlson}, \citenamefont {Gandolfi}, \citenamefont {Rocco},\
  and\ \citenamefont {Schiavilla}}]{PhysRevX.10.031068}%
  \BibitemOpen
  \bibfield  {author} {\bibinfo {author} {\bibfnamefont {A.}~\bibnamefont
  {Lovato}}, \bibinfo {author} {\bibfnamefont {J.}~\bibnamefont {Carlson}},
  \bibinfo {author} {\bibfnamefont {S.}~\bibnamefont {Gandolfi}}, \bibinfo
  {author} {\bibfnamefont {N.}~\bibnamefont {Rocco}}, \ and\ \bibinfo {author}
  {\bibfnamefont {R.}~\bibnamefont {Schiavilla}},\ }\href {\doibase
  10.1103/PhysRevX.10.031068} {\bibfield  {journal} {\bibinfo  {journal} {Phys.
  Rev. X}\ }\textbf {\bibinfo {volume} {10}},\ \bibinfo {pages} {031068}
  (\bibinfo {year} {2020})}\BibitemShut {NoStop}%
\bibitem [{\citenamefont {Nikolakopoulos}\ \emph {et~al.}(2024)\citenamefont
  {Nikolakopoulos}, \citenamefont {Lovato},\ and\ \citenamefont
  {Rocco}}]{Nikolakopoulos:2023zse}%
  \BibitemOpen
  \bibfield  {author} {\bibinfo {author} {\bibfnamefont {A.}~\bibnamefont
  {Nikolakopoulos}}, \bibinfo {author} {\bibfnamefont {A.}~\bibnamefont
  {Lovato}}, \ and\ \bibinfo {author} {\bibfnamefont {N.}~\bibnamefont
  {Rocco}},\ }\href {\doibase 10.1103/PhysRevC.109.014623} {\bibfield
  {journal} {\bibinfo  {journal} {Phys. Rev. C}\ }\textbf {\bibinfo {volume}
  {109}},\ \bibinfo {pages} {014623} (\bibinfo {year} {2024})},\ \Eprint
  {http://arxiv.org/abs/2304.11772} {arXiv:2304.11772 [nucl-th]} \BibitemShut
  {NoStop}%
\bibitem [{\citenamefont {Sobczyk}\ \emph {et~al.}(2021)\citenamefont
  {Sobczyk}, \citenamefont {Acharya}, \citenamefont {Bacca},\ and\
  \citenamefont {Hagen}}]{Sobczyk:2021dwm}%
  \BibitemOpen
  \bibfield  {author} {\bibinfo {author} {\bibfnamefont {J.~E.}\ \bibnamefont
  {Sobczyk}}, \bibinfo {author} {\bibfnamefont {B.}~\bibnamefont {Acharya}},
  \bibinfo {author} {\bibfnamefont {S.}~\bibnamefont {Bacca}}, \ and\ \bibinfo
  {author} {\bibfnamefont {G.}~\bibnamefont {Hagen}},\ }\href {\doibase
  10.1103/PhysRevLett.127.072501} {\bibfield  {journal} {\bibinfo  {journal}
  {Phys. Rev. Lett.}\ }\textbf {\bibinfo {volume} {127}},\ \bibinfo {pages}
  {072501} (\bibinfo {year} {2021})},\ \Eprint
  {http://arxiv.org/abs/2103.06786} {arXiv:2103.06786 [nucl-th]} \BibitemShut
  {NoStop}%
\bibitem [{\citenamefont {Sobczyk}\ \emph {et~al.}(2024)\citenamefont
  {Sobczyk}, \citenamefont {Acharya}, \citenamefont {Bacca},\ and\
  \citenamefont {Hagen}}]{Sobczyk:2023sxh}%
  \BibitemOpen
  \bibfield  {author} {\bibinfo {author} {\bibfnamefont {J.~E.}\ \bibnamefont
  {Sobczyk}}, \bibinfo {author} {\bibfnamefont {B.}~\bibnamefont {Acharya}},
  \bibinfo {author} {\bibfnamefont {S.}~\bibnamefont {Bacca}}, \ and\ \bibinfo
  {author} {\bibfnamefont {G.}~\bibnamefont {Hagen}},\ }\href {\doibase
  10.1103/PhysRevC.109.025502} {\bibfield  {journal} {\bibinfo  {journal}
  {Phys. Rev. C}\ }\textbf {\bibinfo {volume} {109}},\ \bibinfo {pages}
  {025502} (\bibinfo {year} {2024})},\ \Eprint
  {http://arxiv.org/abs/2310.03109} {arXiv:2310.03109 [nucl-th]} \BibitemShut
  {NoStop}%
\bibitem [{\citenamefont {Acharya}\ \emph {et~al.}(2025)\citenamefont
  {Acharya}, \citenamefont {Sobczyk}, \citenamefont {Bacca}, \citenamefont
  {Hagen},\ and\ \citenamefont {Jiang}}]{Acharya:2024xah}%
  \BibitemOpen
  \bibfield  {author} {\bibinfo {author} {\bibfnamefont {B.}~\bibnamefont
  {Acharya}}, \bibinfo {author} {\bibfnamefont {J.~E.}\ \bibnamefont
  {Sobczyk}}, \bibinfo {author} {\bibfnamefont {S.}~\bibnamefont {Bacca}},
  \bibinfo {author} {\bibfnamefont {G.}~\bibnamefont {Hagen}}, \ and\ \bibinfo
  {author} {\bibfnamefont {W.}~\bibnamefont {Jiang}},\ }\href {\doibase
  10.1103/PhysRevLett.134.202501} {\bibfield  {journal} {\bibinfo  {journal}
  {Phys. Rev. Lett.}\ }\textbf {\bibinfo {volume} {134}},\ \bibinfo {pages}
  {202501} (\bibinfo {year} {2025})},\ \Eprint
  {http://arxiv.org/abs/2410.05962} {arXiv:2410.05962 [nucl-th]} \BibitemShut
  {NoStop}%
\bibitem [{\citenamefont {Miorelli}\ \emph
  {et~al.}(2016{\natexlab{a}})\citenamefont {Miorelli}, \citenamefont {Bacca},
  \citenamefont {Barnea}, \citenamefont {Hagen}, \citenamefont {Jansen},
  \citenamefont {Orlandini},\ and\ \citenamefont
  {Papenbrock}}]{Miorelli:2016qbk}%
  \BibitemOpen
  \bibfield  {author} {\bibinfo {author} {\bibfnamefont {M.}~\bibnamefont
  {Miorelli}}, \bibinfo {author} {\bibfnamefont {S.}~\bibnamefont {Bacca}},
  \bibinfo {author} {\bibfnamefont {N.}~\bibnamefont {Barnea}}, \bibinfo
  {author} {\bibfnamefont {G.}~\bibnamefont {Hagen}}, \bibinfo {author}
  {\bibfnamefont {G.~R.}\ \bibnamefont {Jansen}}, \bibinfo {author}
  {\bibfnamefont {G.}~\bibnamefont {Orlandini}}, \ and\ \bibinfo {author}
  {\bibfnamefont {T.}~\bibnamefont {Papenbrock}},\ }\href {\doibase
  10.1103/PhysRevC.94.034317} {\bibfield  {journal} {\bibinfo  {journal} {Phys.
  Rev. C}\ }\textbf {\bibinfo {volume} {94}},\ \bibinfo {pages} {034317}
  (\bibinfo {year} {2016}{\natexlab{a}})},\ \Eprint
  {http://arxiv.org/abs/1604.05381} {arXiv:1604.05381 [nucl-th]} \BibitemShut
  {NoStop}%
\bibitem [{\citenamefont {Birkhan}\ \emph {et~al.}(2017)\citenamefont {Birkhan}
  \emph {et~al.}}]{Birkhan:2016qkr}%
  \BibitemOpen
  \bibfield  {author} {\bibinfo {author} {\bibfnamefont {J.}~\bibnamefont
  {Birkhan}} \emph {et~al.},\ }\href {\doibase 10.1103/PhysRevLett.118.252501}
  {\bibfield  {journal} {\bibinfo  {journal} {Phys. Rev. Lett.}\ }\textbf
  {\bibinfo {volume} {118}},\ \bibinfo {pages} {252501} (\bibinfo {year}
  {2017})},\ \Eprint {http://arxiv.org/abs/1611.07072} {arXiv:1611.07072
  [nucl-ex]} \BibitemShut {NoStop}%
\bibitem [{\citenamefont {Lovato}\ \emph {et~al.}(2014)\citenamefont {Lovato},
  \citenamefont {Gandolfi}, \citenamefont {Carlson}, \citenamefont {Pieper},\
  and\ \citenamefont {Schiavilla}}]{Lovato14}%
  \BibitemOpen
  \bibfield  {author} {\bibinfo {author} {\bibfnamefont {A.}~\bibnamefont
  {Lovato}}, \bibinfo {author} {\bibfnamefont {S.}~\bibnamefont {Gandolfi}},
  \bibinfo {author} {\bibfnamefont {J.}~\bibnamefont {Carlson}}, \bibinfo
  {author} {\bibfnamefont {S.}~\bibnamefont {Pieper}}, \ and\ \bibinfo {author}
  {\bibfnamefont {R.}~\bibnamefont {Schiavilla}},\ }\href@noop {} {\bibfield
  {journal} {\bibinfo  {journal} {Phys. Rev. Lett.}\ }\textbf {\bibinfo
  {volume} {112}},\ \bibinfo {pages} {182502} (\bibinfo {year}
  {2014})}\BibitemShut {NoStop}%
\bibitem [{\citenamefont {Carlson}\ \emph {et~al.}(2002)\citenamefont
  {Carlson}, \citenamefont {Jourdan}, \citenamefont {Schiavilla},\ and\
  \citenamefont {Sick}}]{PhysRevC.65.024002}%
  \BibitemOpen
  \bibfield  {author} {\bibinfo {author} {\bibfnamefont {J.}~\bibnamefont
  {Carlson}}, \bibinfo {author} {\bibfnamefont {J.}~\bibnamefont {Jourdan}},
  \bibinfo {author} {\bibfnamefont {R.}~\bibnamefont {Schiavilla}}, \ and\
  \bibinfo {author} {\bibfnamefont {I.}~\bibnamefont {Sick}},\ }\href {\doibase
  10.1103/PhysRevC.65.024002} {\bibfield  {journal} {\bibinfo  {journal} {Phys.
  Rev. C}\ }\textbf {\bibinfo {volume} {65}},\ \bibinfo {pages} {024002}
  (\bibinfo {year} {2002})}\BibitemShut {NoStop}%
\bibitem [{\citenamefont {Carlson}\ and\ \citenamefont
  {Schiavilla}(1994)}]{PhysRevC.49.R2880}%
  \BibitemOpen
  \bibfield  {author} {\bibinfo {author} {\bibfnamefont {J.}~\bibnamefont
  {Carlson}}\ and\ \bibinfo {author} {\bibfnamefont {R.}~\bibnamefont
  {Schiavilla}},\ }\href {\doibase 10.1103/PhysRevC.49.R2880} {\bibfield
  {journal} {\bibinfo  {journal} {Phys. Rev. C}\ }\textbf {\bibinfo {volume}
  {49}},\ \bibinfo {pages} {R2880} (\bibinfo {year} {1994})}\BibitemShut
  {NoStop}%
\bibitem [{\citenamefont {Hagen}\ \emph {et~al.}(2014)\citenamefont {Hagen},
  \citenamefont {Papenbrock}, \citenamefont {Hjorth-Jensen},\ and\
  \citenamefont {Dean}}]{Hagen:2013nca}%
  \BibitemOpen
  \bibfield  {author} {\bibinfo {author} {\bibfnamefont {G.}~\bibnamefont
  {Hagen}}, \bibinfo {author} {\bibfnamefont {T.}~\bibnamefont {Papenbrock}},
  \bibinfo {author} {\bibfnamefont {M.}~\bibnamefont {Hjorth-Jensen}}, \ and\
  \bibinfo {author} {\bibfnamefont {D.~J.}\ \bibnamefont {Dean}},\ }\href
  {\doibase 10.1088/0034-4885/77/9/096302} {\bibfield  {journal} {\bibinfo
  {journal} {Rept. Prog. Phys.}\ }\textbf {\bibinfo {volume} {77}},\ \bibinfo
  {pages} {096302} (\bibinfo {year} {2014})},\ \Eprint
  {http://arxiv.org/abs/1312.7872} {arXiv:1312.7872 [nucl-th]} \BibitemShut
  {NoStop}%
\bibitem [{\citenamefont {Barnea}\ \emph {et~al.}(2000)\citenamefont {Barnea},
  \citenamefont {Leidemann},\ and\ \citenamefont {Orlandini}}]{Barnea:1999be}%
  \BibitemOpen
  \bibfield  {author} {\bibinfo {author} {\bibfnamefont {N.}~\bibnamefont
  {Barnea}}, \bibinfo {author} {\bibfnamefont {W.}~\bibnamefont {Leidemann}}, \
  and\ \bibinfo {author} {\bibfnamefont {G.}~\bibnamefont {Orlandini}},\ }\href
  {\doibase 10.1103/PhysRevC.61.054001} {\bibfield  {journal} {\bibinfo
  {journal} {Phys. Rev. C}\ }\textbf {\bibinfo {volume} {61}},\ \bibinfo
  {pages} {054001} (\bibinfo {year} {2000})},\ \Eprint
  {http://arxiv.org/abs/nucl-th/9910062} {arXiv:nucl-th/9910062} \BibitemShut
  {NoStop}%
\bibitem [{\citenamefont {Efros}\ \emph {et~al.}(1994)\citenamefont {Efros},
  \citenamefont {Leidemann},\ and\ \citenamefont {Orlandini}}]{EFROS1994130}%
  \BibitemOpen
  \bibfield  {author} {\bibinfo {author} {\bibfnamefont {V.~D.}\ \bibnamefont
  {Efros}}, \bibinfo {author} {\bibfnamefont {W.}~\bibnamefont {Leidemann}}, \
  and\ \bibinfo {author} {\bibfnamefont {G.}~\bibnamefont {Orlandini}},\ }\href
  {\doibase https://doi.org/10.1016/0370-2693(94)91355-2} {\bibfield  {journal}
  {\bibinfo  {journal} {Physics Letters B}\ }\textbf {\bibinfo {volume}
  {338}},\ \bibinfo {pages} {130} (\bibinfo {year} {1994})}\BibitemShut
  {NoStop}%
\bibitem [{\citenamefont {Lovato}\ \emph {et~al.}(2018)\citenamefont {Lovato},
  \citenamefont {Gandolfi}, \citenamefont {Carlson}, \citenamefont {Lusk},
  \citenamefont {Pieper},\ and\ \citenamefont {Schiavilla}}]{Lovato:2017cux}%
  \BibitemOpen
  \bibfield  {author} {\bibinfo {author} {\bibfnamefont {A.}~\bibnamefont
  {Lovato}}, \bibinfo {author} {\bibfnamefont {S.}~\bibnamefont {Gandolfi}},
  \bibinfo {author} {\bibfnamefont {J.}~\bibnamefont {Carlson}}, \bibinfo
  {author} {\bibfnamefont {E.}~\bibnamefont {Lusk}}, \bibinfo {author}
  {\bibfnamefont {S.~C.}\ \bibnamefont {Pieper}}, \ and\ \bibinfo {author}
  {\bibfnamefont {R.}~\bibnamefont {Schiavilla}},\ }\href {\doibase
  10.1103/PhysRevC.97.022502} {\bibfield  {journal} {\bibinfo  {journal} {Phys.
  Rev. C}\ }\textbf {\bibinfo {volume} {97}},\ \bibinfo {pages} {022502}
  (\bibinfo {year} {2018})},\ \Eprint {http://arxiv.org/abs/1711.02047}
  {arXiv:1711.02047 [nucl-th]} \BibitemShut {NoStop}%
\bibitem [{\citenamefont {Lovato}\ \emph {et~al.}(2015)\citenamefont {Lovato},
  \citenamefont {Gandolfi}, \citenamefont {Carlson}, \citenamefont {Pieper},\
  and\ \citenamefont {Schiavilla}}]{PhysRevC.91.062501}%
  \BibitemOpen
  \bibfield  {author} {\bibinfo {author} {\bibfnamefont {A.}~\bibnamefont
  {Lovato}}, \bibinfo {author} {\bibfnamefont {S.}~\bibnamefont {Gandolfi}},
  \bibinfo {author} {\bibfnamefont {J.}~\bibnamefont {Carlson}}, \bibinfo
  {author} {\bibfnamefont {S.~C.}\ \bibnamefont {Pieper}}, \ and\ \bibinfo
  {author} {\bibfnamefont {R.}~\bibnamefont {Schiavilla}},\ }\href {\doibase
  10.1103/PhysRevC.91.062501} {\bibfield  {journal} {\bibinfo  {journal} {Phys.
  Rev. C}\ }\textbf {\bibinfo {volume} {91}},\ \bibinfo {pages} {062501}
  (\bibinfo {year} {2015})}\BibitemShut {NoStop}%
\bibitem [{\citenamefont {Raghavan}\ \emph {et~al.}(2021)\citenamefont
  {Raghavan}, \citenamefont {Balaprakash}, \citenamefont {Lovato},
  \citenamefont {Rocco},\ and\ \citenamefont {Wild}}]{raghavan_machine_2021}%
  \BibitemOpen
  \bibfield  {author} {\bibinfo {author} {\bibfnamefont {K.}~\bibnamefont
  {Raghavan}}, \bibinfo {author} {\bibfnamefont {P.}~\bibnamefont
  {Balaprakash}}, \bibinfo {author} {\bibfnamefont {A.}~\bibnamefont {Lovato}},
  \bibinfo {author} {\bibfnamefont {N.}~\bibnamefont {Rocco}}, \ and\ \bibinfo
  {author} {\bibfnamefont {S.~M.}\ \bibnamefont {Wild}},\ }\href {\doibase
  10.1103/PhysRevC.103.035502} {\bibfield  {journal} {\bibinfo  {journal}
  {Phys. Rev. C}\ }\textbf {\bibinfo {volume} {103}},\ \bibinfo {pages}
  {035502} (\bibinfo {year} {2021})}\BibitemShut {NoStop}%
\bibitem [{\citenamefont {Raghavan}\ and\ \citenamefont
  {Lovato}(2024)}]{raghavan_uncertainty-quantification-enabled_2024}%
  \BibitemOpen
  \bibfield  {author} {\bibinfo {author} {\bibfnamefont {K.}~\bibnamefont
  {Raghavan}}\ and\ \bibinfo {author} {\bibfnamefont {A.}~\bibnamefont
  {Lovato}},\ }\href {\doibase 10.1103/PhysRevC.110.025504} {\bibfield
  {journal} {\bibinfo  {journal} {Phys. Rev. C}\ }\textbf {\bibinfo {volume}
  {110}},\ \bibinfo {pages} {025504} (\bibinfo {year} {2024})}\BibitemShut
  {NoStop}%
\bibitem [{\citenamefont {Sobczyk}\ and\ \citenamefont
  {Roggero}(2022)}]{Sobczyk:2021ejs}%
  \BibitemOpen
  \bibfield  {author} {\bibinfo {author} {\bibfnamefont {J.~E.}\ \bibnamefont
  {Sobczyk}}\ and\ \bibinfo {author} {\bibfnamefont {A.}~\bibnamefont
  {Roggero}},\ }\href {\doibase 10.1103/PhysRevE.105.055310} {\bibfield
  {journal} {\bibinfo  {journal} {Phys. Rev. E}\ }\textbf {\bibinfo {volume}
  {105}},\ \bibinfo {pages} {055310} (\bibinfo {year} {2022})},\ \Eprint
  {http://arxiv.org/abs/2110.02108} {arXiv:2110.02108 [nucl-th]} \BibitemShut
  {NoStop}%
\bibitem [{\citenamefont {Parnes}\ \emph {et~al.}(2026)\citenamefont {Parnes},
  \citenamefont {Barnea}, \citenamefont {Carleo}, \citenamefont {Lovato},
  \citenamefont {Rocco},\ and\ \citenamefont {Zhang}}]{Parnes:2025seu}%
  \BibitemOpen
  \bibfield  {author} {\bibinfo {author} {\bibfnamefont {E.}~\bibnamefont
  {Parnes}}, \bibinfo {author} {\bibfnamefont {N.}~\bibnamefont {Barnea}},
  \bibinfo {author} {\bibfnamefont {G.}~\bibnamefont {Carleo}}, \bibinfo
  {author} {\bibfnamefont {A.}~\bibnamefont {Lovato}}, \bibinfo {author}
  {\bibfnamefont {N.}~\bibnamefont {Rocco}}, \ and\ \bibinfo {author}
  {\bibfnamefont {X.}~\bibnamefont {Zhang}},\ }\href {\doibase
  10.1103/tlqz-nw28} {\bibfield  {journal} {\bibinfo  {journal} {Phys. Rev.
  Lett.}\ }\textbf {\bibinfo {volume} {136}},\ \bibinfo {pages} {032501}
  (\bibinfo {year} {2026})},\ \Eprint {http://arxiv.org/abs/2504.20195}
  {arXiv:2504.20195 [nucl-th]} \BibitemShut {NoStop}%
\bibitem [{\citenamefont {Jay}(2025)}]{Jay:2025dzl}%
  \BibitemOpen
  \bibfield  {author} {\bibinfo {author} {\bibfnamefont {W.}~\bibnamefont
  {Jay}},\ }in\ \href@noop {} {\emph {\bibinfo {booktitle} {{41st International
  Symposium on Lattice Field Theory}}}}\ (\bibinfo {year} {2025})\ \Eprint
  {http://arxiv.org/abs/2501.12259} {arXiv:2501.12259 [hep-lat]} \BibitemShut
  {NoStop}%
\bibitem [{\citenamefont {Salg}\ \emph {et~al.}(2025)\citenamefont {Salg},
  \citenamefont {Romero-L{\'o}pez},\ and\ \citenamefont {Jay}}]{Salg:2025now}%
  \BibitemOpen
  \bibfield  {author} {\bibinfo {author} {\bibfnamefont {M.}~\bibnamefont
  {Salg}}, \bibinfo {author} {\bibfnamefont {F.}~\bibnamefont
  {Romero-L{\'o}pez}}, \ and\ \bibinfo {author} {\bibfnamefont {W.~I.}\
  \bibnamefont {Jay}},\ }\href {\doibase 10.1103/ty19-xvvw} {\bibfield
  {journal} {\bibinfo  {journal} {Phys. Rev. D}\ }\textbf {\bibinfo {volume}
  {112}},\ \bibinfo {pages} {114502} (\bibinfo {year} {2025})},\ \Eprint
  {http://arxiv.org/abs/2506.16161} {arXiv:2506.16161 [hep-lat]} \BibitemShut
  {NoStop}%
\bibitem [{\citenamefont {Walecka}(2004)}]{walecka04}%
  \BibitemOpen
  \bibfield  {author} {\bibinfo {author} {\bibfnamefont {J.}~\bibnamefont
  {Walecka}},\ }\href {https://books.google.be/books?id=sBABwQEACAAJ} {\emph
  {\bibinfo {title} {Theoretical Nuclear and Subnuclear Physics}}}\ (\bibinfo
  {publisher} {World Scientfic, Imperial College Press},\ \bibinfo {year}
  {2004})\BibitemShut {NoStop}%
\bibitem [{\citenamefont {Aguilar-Arevalo}\ \emph {et~al.}(2009)\citenamefont
  {Aguilar-Arevalo}, \citenamefont {Anderson}, \citenamefont {Bazarko},
  \citenamefont {Brice}, \citenamefont {Brown}, \citenamefont {Bugel} \emph
  {et~al.}}]{MBflux:2009}%
  \BibitemOpen
  \bibfield  {author} {\bibinfo {author} {\bibfnamefont {A.~A.}\ \bibnamefont
  {Aguilar-Arevalo}}, \bibinfo {author} {\bibfnamefont {C.~E.}\ \bibnamefont
  {Anderson}}, \bibinfo {author} {\bibfnamefont {A.~O.}\ \bibnamefont
  {Bazarko}}, \bibinfo {author} {\bibfnamefont {S.~J.}\ \bibnamefont {Brice}},
  \bibinfo {author} {\bibfnamefont {B.~C.}\ \bibnamefont {Brown}}, \bibinfo
  {author} {\bibfnamefont {L.}~\bibnamefont {Bugel}},  \emph {et~al.} (\bibinfo
  {collaboration} {MiniBooNE Collaboration}),\ }\href {\doibase
  10.1103/PhysRevD.79.072002} {\bibfield  {journal} {\bibinfo  {journal} {Phys.
  Rev. D}\ }\textbf {\bibinfo {volume} {79}},\ \bibinfo {pages} {072002}
  (\bibinfo {year} {2009})}\BibitemShut {NoStop}%
\bibitem [{\citenamefont {Lovato}\ \emph {et~al.}(2025)\citenamefont {Lovato},
  \citenamefont {Rocco},\ and\ \citenamefont {Steinberg}}]{Lovato:2023khk}%
  \BibitemOpen
  \bibfield  {author} {\bibinfo {author} {\bibfnamefont {A.}~\bibnamefont
  {Lovato}}, \bibinfo {author} {\bibfnamefont {N.}~\bibnamefont {Rocco}}, \
  and\ \bibinfo {author} {\bibfnamefont {N.}~\bibnamefont {Steinberg}},\ }\href
  {\doibase 10.1103/m645-5whh} {\bibfield  {journal} {\bibinfo  {journal}
  {Phys. Rev. C}\ }\textbf {\bibinfo {volume} {112}},\ \bibinfo {pages}
  {045501} (\bibinfo {year} {2025})},\ \Eprint
  {http://arxiv.org/abs/2312.12545} {arXiv:2312.12545 [nucl-th]} \BibitemShut
  {NoStop}%
\bibitem [{\citenamefont {Korover}\ \emph {et~al.}(2023)\citenamefont {Korover}
  \emph {et~al.}}]{CLAS:2022odn}%
  \BibitemOpen
  \bibfield  {author} {\bibinfo {author} {\bibfnamefont {I.}~\bibnamefont
  {Korover}} \emph {et~al.} (\bibinfo {collaboration} {CLAS}),\ }\href
  {\doibase 10.1103/PhysRevC.107.L061301} {\bibfield  {journal} {\bibinfo
  {journal} {Phys. Rev. C}\ }\textbf {\bibinfo {volume} {107}},\ \bibinfo
  {pages} {L061301} (\bibinfo {year} {2023})},\ \Eprint
  {http://arxiv.org/abs/2209.01492} {arXiv:2209.01492 [nucl-ex]} \BibitemShut
  {NoStop}%
\bibitem [{VMC two-body momentum distributions()}]{nkk_web}%
  \BibitemOpen
  VMC two-body momentum distributions,\ \href@noop {} {\enquote {\bibinfo
  {title} {Quantum monte carlo results for two-nucleon momentum
  distributions},}\ }\bibinfo {howpublished}
  {\url{https://www.phy.anl.gov/theory/research/momenta2/}}\BibitemShut
  {NoStop}%
\bibitem [{\citenamefont {Tropiano}\ \emph {et~al.}(2021)\citenamefont
  {Tropiano}, \citenamefont {Bogner},\ and\ \citenamefont
  {Furnstahl}}]{Tropiano:2021qgf}%
  \BibitemOpen
  \bibfield  {author} {\bibinfo {author} {\bibfnamefont {A.~J.}\ \bibnamefont
  {Tropiano}}, \bibinfo {author} {\bibfnamefont {S.~K.}\ \bibnamefont
  {Bogner}}, \ and\ \bibinfo {author} {\bibfnamefont {R.~J.}\ \bibnamefont
  {Furnstahl}},\ }\href {\doibase 10.1103/PhysRevC.104.034311} {\bibfield
  {journal} {\bibinfo  {journal} {Phys. Rev. C}\ }\textbf {\bibinfo {volume}
  {104}},\ \bibinfo {pages} {034311} (\bibinfo {year} {2021})},\ \Eprint
  {http://arxiv.org/abs/2105.13936} {arXiv:2105.13936 [nucl-th]} \BibitemShut
  {NoStop}%
\bibitem [{\citenamefont {Ryckebusch}\ \emph {et~al.}(2019)\citenamefont
  {Ryckebusch}, \citenamefont {Cosyn}, \citenamefont {Vieijra},\ and\
  \citenamefont {Casert}}]{Ryckebusch:2019oya}%
  \BibitemOpen
  \bibfield  {author} {\bibinfo {author} {\bibfnamefont {J.}~\bibnamefont
  {Ryckebusch}}, \bibinfo {author} {\bibfnamefont {W.}~\bibnamefont {Cosyn}},
  \bibinfo {author} {\bibfnamefont {T.}~\bibnamefont {Vieijra}}, \ and\
  \bibinfo {author} {\bibfnamefont {C.}~\bibnamefont {Casert}},\ }\href
  {\doibase 10.1103/PhysRevC.100.054620} {\bibfield  {journal} {\bibinfo
  {journal} {Phys. Rev. C}\ }\textbf {\bibinfo {volume} {100}},\ \bibinfo
  {pages} {054620} (\bibinfo {year} {2019})},\ \Eprint
  {http://arxiv.org/abs/1907.07259} {arXiv:1907.07259 [nucl-th]} \BibitemShut
  {NoStop}%
\bibitem [{\citenamefont {Cruz-Torres}\ \emph {et~al.}(2021)\citenamefont
  {Cruz-Torres}, \citenamefont {Lonardoni}, \citenamefont {Weiss},
  \citenamefont {Barnea}, \citenamefont {Higinbotham}, \citenamefont
  {Piasetzky}, \citenamefont {Schmidt}, \citenamefont {Weinstein},
  \citenamefont {Wiringa},\ and\ \citenamefont {Hen}}]{Cruz-Torres:2019fum}%
  \BibitemOpen
  \bibfield  {author} {\bibinfo {author} {\bibfnamefont {R.}~\bibnamefont
  {Cruz-Torres}}, \bibinfo {author} {\bibfnamefont {D.}~\bibnamefont
  {Lonardoni}}, \bibinfo {author} {\bibfnamefont {R.}~\bibnamefont {Weiss}},
  \bibinfo {author} {\bibfnamefont {N.}~\bibnamefont {Barnea}}, \bibinfo
  {author} {\bibfnamefont {D.~W.}\ \bibnamefont {Higinbotham}}, \bibinfo
  {author} {\bibfnamefont {E.}~\bibnamefont {Piasetzky}}, \bibinfo {author}
  {\bibfnamefont {A.}~\bibnamefont {Schmidt}}, \bibinfo {author} {\bibfnamefont
  {L.~B.}\ \bibnamefont {Weinstein}}, \bibinfo {author} {\bibfnamefont {R.~B.}\
  \bibnamefont {Wiringa}}, \ and\ \bibinfo {author} {\bibfnamefont
  {O.}~\bibnamefont {Hen}},\ }\href {\doibase 10.1038/s41567-020-01053-7}
  {\bibfield  {journal} {\bibinfo  {journal} {Nature Phys.}\ }\textbf {\bibinfo
  {volume} {17}},\ \bibinfo {pages} {306} (\bibinfo {year} {2021})},\ \Eprint
  {http://arxiv.org/abs/1907.03658} {arXiv:1907.03658 [nucl-th]} \BibitemShut
  {NoStop}%
\bibitem [{\citenamefont {Ciofi~degli Atti}(2015)}]{CiofidegliAtti:2015lcu}%
  \BibitemOpen
  \bibfield  {author} {\bibinfo {author} {\bibfnamefont {C.}~\bibnamefont
  {Ciofi~degli Atti}},\ }\href {\doibase 10.1016/j.physrep.2015.06.002}
  {\bibfield  {journal} {\bibinfo  {journal} {Phys. Rept.}\ }\textbf {\bibinfo
  {volume} {590}},\ \bibinfo {pages} {1} (\bibinfo {year} {2015})}\BibitemShut
  {NoStop}%
\bibitem [{\citenamefont {Ryckebusch}\ \emph {et~al.}(2015)\citenamefont
  {Ryckebusch}, \citenamefont {Cosyn},\ and\ \citenamefont
  {Vanhalst}}]{Ryckebusch:2014ann}%
  \BibitemOpen
  \bibfield  {author} {\bibinfo {author} {\bibfnamefont {J.}~\bibnamefont
  {Ryckebusch}}, \bibinfo {author} {\bibfnamefont {W.}~\bibnamefont {Cosyn}}, \
  and\ \bibinfo {author} {\bibfnamefont {M.}~\bibnamefont {Vanhalst}},\ }\href
  {\doibase 10.1088/0954-3899/42/5/055104} {\bibfield  {journal} {\bibinfo
  {journal} {J. Phys. G}\ }\textbf {\bibinfo {volume} {42}},\ \bibinfo {pages}
  {055104} (\bibinfo {year} {2015})},\ \Eprint {http://arxiv.org/abs/1405.3814}
  {arXiv:1405.3814 [nucl-th]} \BibitemShut {NoStop}%
\bibitem [{\citenamefont {Moreno}\ \emph {et~al.}(2014)\citenamefont {Moreno},
  \citenamefont {Donnelly}, \citenamefont {Van~Orden},\ and\ \citenamefont
  {Ford}}]{Moreno:2014kia}%
  \BibitemOpen
  \bibfield  {author} {\bibinfo {author} {\bibfnamefont {O.}~\bibnamefont
  {Moreno}}, \bibinfo {author} {\bibfnamefont {T.~W.}\ \bibnamefont
  {Donnelly}}, \bibinfo {author} {\bibfnamefont {J.~W.}\ \bibnamefont
  {Van~Orden}}, \ and\ \bibinfo {author} {\bibfnamefont {W.~P.}\ \bibnamefont
  {Ford}},\ }\href {\doibase 10.1103/PhysRevD.90.013014} {\bibfield  {journal}
  {\bibinfo  {journal} {Phys. Rev. D}\ }\textbf {\bibinfo {volume} {90}},\
  \bibinfo {pages} {013014} (\bibinfo {year} {2014})},\ \Eprint
  {http://arxiv.org/abs/1406.4494} {arXiv:1406.4494 [hep-th]} \BibitemShut
  {NoStop}%
\bibitem [{\citenamefont {Nevo~Dinur}\ \emph {et~al.}(2014)\citenamefont
  {Nevo~Dinur}, \citenamefont {Barnea}, \citenamefont {Ji},\ and\ \citenamefont
  {Bacca}}]{PhysRevC.89.064317}%
  \BibitemOpen
  \bibfield  {author} {\bibinfo {author} {\bibfnamefont {N.}~\bibnamefont
  {Nevo~Dinur}}, \bibinfo {author} {\bibfnamefont {N.}~\bibnamefont {Barnea}},
  \bibinfo {author} {\bibfnamefont {C.}~\bibnamefont {Ji}}, \ and\ \bibinfo
  {author} {\bibfnamefont {S.}~\bibnamefont {Bacca}},\ }\href {\doibase
  10.1103/PhysRevC.89.064317} {\bibfield  {journal} {\bibinfo  {journal} {Phys.
  Rev. C}\ }\textbf {\bibinfo {volume} {89}},\ \bibinfo {pages} {064317}
  (\bibinfo {year} {2014})}\BibitemShut {NoStop}%
\bibitem [{\citenamefont {Miorelli}\ \emph
  {et~al.}(2016{\natexlab{b}})\citenamefont {Miorelli}, \citenamefont {Bacca},
  \citenamefont {Barnea}, \citenamefont {Hagen}, \citenamefont {Jansen},
  \citenamefont {Orlandini},\ and\ \citenamefont
  {Papenbrock}}]{PhysRevC.94.034317}%
  \BibitemOpen
  \bibfield  {author} {\bibinfo {author} {\bibfnamefont {M.}~\bibnamefont
  {Miorelli}}, \bibinfo {author} {\bibfnamefont {S.}~\bibnamefont {Bacca}},
  \bibinfo {author} {\bibfnamefont {N.}~\bibnamefont {Barnea}}, \bibinfo
  {author} {\bibfnamefont {G.}~\bibnamefont {Hagen}}, \bibinfo {author}
  {\bibfnamefont {G.~R.}\ \bibnamefont {Jansen}}, \bibinfo {author}
  {\bibfnamefont {G.}~\bibnamefont {Orlandini}}, \ and\ \bibinfo {author}
  {\bibfnamefont {T.}~\bibnamefont {Papenbrock}},\ }\href {\doibase
  10.1103/PhysRevC.94.034317} {\bibfield  {journal} {\bibinfo  {journal} {Phys.
  Rev. C}\ }\textbf {\bibinfo {volume} {94}},\ \bibinfo {pages} {034317}
  (\bibinfo {year} {2016}{\natexlab{b}})}\BibitemShut {NoStop}%
\bibitem [{\citenamefont {Bonaiti}\ \emph {et~al.}(2022)\citenamefont
  {Bonaiti}, \citenamefont {Bacca},\ and\ \citenamefont
  {Hagen}}]{PhysRevC.105.034313}%
  \BibitemOpen
  \bibfield  {author} {\bibinfo {author} {\bibfnamefont {F.}~\bibnamefont
  {Bonaiti}}, \bibinfo {author} {\bibfnamefont {S.}~\bibnamefont {Bacca}}, \
  and\ \bibinfo {author} {\bibfnamefont {G.}~\bibnamefont {Hagen}},\ }\href
  {\doibase 10.1103/PhysRevC.105.034313} {\bibfield  {journal} {\bibinfo
  {journal} {Phys. Rev. C}\ }\textbf {\bibinfo {volume} {105}},\ \bibinfo
  {pages} {034313} (\bibinfo {year} {2022})}\BibitemShut {NoStop}%
\bibitem [{\citenamefont {Bacca}()}]{Baccatalk}%
  \BibitemOpen
  \bibfield  {author} {\bibinfo {author} {\bibfnamefont {S.}~\bibnamefont
  {Bacca}},\ }\href@noop {} {\ }\bibinfo {note} {(private
  communication)}\BibitemShut {NoStop}%
\bibitem [{\citenamefont {Meyer}\ \emph {et~al.}(2016)\citenamefont {Meyer},
  \citenamefont {Betancourt}, \citenamefont {Gran},\ and\ \citenamefont
  {Hill}}]{Meyer:2016oeg}%
  \BibitemOpen
  \bibfield  {author} {\bibinfo {author} {\bibfnamefont {A.~S.}\ \bibnamefont
  {Meyer}}, \bibinfo {author} {\bibfnamefont {M.}~\bibnamefont {Betancourt}},
  \bibinfo {author} {\bibfnamefont {R.}~\bibnamefont {Gran}}, \ and\ \bibinfo
  {author} {\bibfnamefont {R.~J.}\ \bibnamefont {Hill}},\ }\href {\doibase
  10.1103/PhysRevD.93.113015} {\bibfield  {journal} {\bibinfo  {journal} {Phys.
  Rev. D}\ }\textbf {\bibinfo {volume} {93}},\ \bibinfo {pages} {113015}
  (\bibinfo {year} {2016})},\ \Eprint {http://arxiv.org/abs/1603.03048}
  {arXiv:1603.03048 [hep-ph]} \BibitemShut {NoStop}%
\bibitem [{\citenamefont {Meyer}\ \emph {et~al.}(2022)\citenamefont {Meyer},
  \citenamefont {Walker-Loud},\ and\ \citenamefont
  {Wilkinson}}]{Meyer:2022mix}%
  \BibitemOpen
  \bibfield  {author} {\bibinfo {author} {\bibfnamefont {A.~S.}\ \bibnamefont
  {Meyer}}, \bibinfo {author} {\bibfnamefont {A.}~\bibnamefont {Walker-Loud}},
  \ and\ \bibinfo {author} {\bibfnamefont {C.}~\bibnamefont {Wilkinson}},\
  }\href {\doibase 10.1146/annurev-nucl-010622-120608} {\bibfield  {journal}
  {\bibinfo  {journal} {Ann. Rev. Nucl. Part. Sci.}\ }\textbf {\bibinfo
  {volume} {72}},\ \bibinfo {pages} {205} (\bibinfo {year} {2022})},\ \Eprint
  {http://arxiv.org/abs/2201.01839} {arXiv:2201.01839 [hep-lat]} \BibitemShut
  {NoStop}%
\bibitem [{\citenamefont {Nikolakopoulos}\ \emph {et~al.}()\citenamefont
  {Nikolakopoulos} \emph {et~al.}}]{Nikolakopoulos:prep_EM}%
  \BibitemOpen
  \bibfield  {author} {\bibinfo {author} {\bibfnamefont {A.}~\bibnamefont
  {Nikolakopoulos}} \emph {et~al.},\ }\href@noop {} {\bibinfo  {journal} {(in
  preparation)}\ }\BibitemShut {NoStop}%
\bibitem [{\citenamefont {Donnelly}(1985)}]{Donnelly85}%
  \BibitemOpen
\bibfield  {journal} {  }\bibfield  {author} {\bibinfo {author} {\bibfnamefont
  {T.~W.}\ \bibnamefont {Donnelly}},\ }\href@noop {} {\bibfield  {journal}
  {\bibinfo  {journal} {Prog. Part. Nucl. Phys.}\ }\textbf {\bibinfo {volume}
  {13}},\ \bibinfo {pages} {183 } (\bibinfo {year} {1985})}\BibitemShut
  {NoStop}%
\bibitem [{\citenamefont {Ciofi~degli Atti}\ and\ \citenamefont
  {Simula}(1996)}]{CiofidegliAtti:1995qe}%
  \BibitemOpen
  \bibfield  {author} {\bibinfo {author} {\bibfnamefont {C.}~\bibnamefont
  {Ciofi~degli Atti}}\ and\ \bibinfo {author} {\bibfnamefont {S.}~\bibnamefont
  {Simula}},\ }\href {\doibase 10.1103/PhysRevC.53.1689} {\bibfield  {journal}
  {\bibinfo  {journal} {Phys. Rev. C}\ }\textbf {\bibinfo {volume} {53}},\
  \bibinfo {pages} {1689} (\bibinfo {year} {1996})},\ \Eprint
  {http://arxiv.org/abs/nucl-th/9507024} {arXiv:nucl-th/9507024} \BibitemShut
  {NoStop}%
\bibitem [{\citenamefont {Jiang}\ \emph {et~al.}(2023)\citenamefont {Jiang}
  \emph {et~al.}}]{JeffersonLabHallA:2022ljj}%
  \BibitemOpen
  \bibfield  {author} {\bibinfo {author} {\bibfnamefont {L.}~\bibnamefont
  {Jiang}} \emph {et~al.} (\bibinfo {collaboration} {Jefferson Lab Hall A}),\
  }\href {\doibase 10.1103/PhysRevD.107.012005} {\bibfield  {journal} {\bibinfo
   {journal} {Phys. Rev. D}\ }\textbf {\bibinfo {volume} {107}},\ \bibinfo
  {pages} {012005} (\bibinfo {year} {2023})},\ \Eprint
  {http://arxiv.org/abs/2209.14108} {arXiv:2209.14108 [nucl-ex]} \BibitemShut
  {NoStop}%
\bibitem [{\citenamefont {Jiang}\ \emph {et~al.}(2022)\citenamefont {Jiang}
  \emph {et~al.}}]{JeffersonLabHallA:2022cit}%
  \BibitemOpen
  \bibfield  {author} {\bibinfo {author} {\bibfnamefont {L.}~\bibnamefont
  {Jiang}} \emph {et~al.} (\bibinfo {collaboration} {Jefferson Lab Hall A}),\
  }\href {\doibase 10.1103/PhysRevD.105.112002} {\bibfield  {journal} {\bibinfo
   {journal} {Phys. Rev. D}\ }\textbf {\bibinfo {volume} {105}},\ \bibinfo
  {pages} {112002} (\bibinfo {year} {2022})},\ \Eprint
  {http://arxiv.org/abs/2203.01748} {arXiv:2203.01748 [nucl-ex]} \BibitemShut
  {NoStop}%
\bibitem [{\citenamefont {Wiringa}\ \emph {et~al.}(2014)\citenamefont
  {Wiringa}, \citenamefont {Schiavilla}, \citenamefont {Pieper},\ and\
  \citenamefont {Carlson}}]{PhysRevC.89.024305}%
  \BibitemOpen
  \bibfield  {author} {\bibinfo {author} {\bibfnamefont {R.~B.}\ \bibnamefont
  {Wiringa}}, \bibinfo {author} {\bibfnamefont {R.}~\bibnamefont {Schiavilla}},
  \bibinfo {author} {\bibfnamefont {S.~C.}\ \bibnamefont {Pieper}}, \ and\
  \bibinfo {author} {\bibfnamefont {J.}~\bibnamefont {Carlson}},\ }\href
  {\doibase 10.1103/PhysRevC.89.024305} {\bibfield  {journal} {\bibinfo
  {journal} {Phys. Rev. C}\ }\textbf {\bibinfo {volume} {89}},\ \bibinfo
  {pages} {024305} (\bibinfo {year} {2014})}\BibitemShut {NoStop}%
\bibitem [{\citenamefont {Wiringa}\ \emph {et~al.}(1995)\citenamefont
  {Wiringa}, \citenamefont {Stoks},\ and\ \citenamefont
  {Schiavilla}}]{PhysRevC.51.38}%
  \BibitemOpen
  \bibfield  {author} {\bibinfo {author} {\bibfnamefont {R.~B.}\ \bibnamefont
  {Wiringa}}, \bibinfo {author} {\bibfnamefont {V.~G.~J.}\ \bibnamefont
  {Stoks}}, \ and\ \bibinfo {author} {\bibfnamefont {R.}~\bibnamefont
  {Schiavilla}},\ }\href {\doibase 10.1103/PhysRevC.51.38} {\bibfield
  {journal} {\bibinfo  {journal} {Phys. Rev. C}\ }\textbf {\bibinfo {volume}
  {51}},\ \bibinfo {pages} {38} (\bibinfo {year} {1995})}\BibitemShut {NoStop}%
\bibitem [{\citenamefont {Piarulli}\ \emph {et~al.}(2023)\citenamefont
  {Piarulli}, \citenamefont {Pastore}, \citenamefont {Wiringa}, \citenamefont
  {Brusilow},\ and\ \citenamefont {Lim}}]{PhysRevC.107.014314}%
  \BibitemOpen
  \bibfield  {author} {\bibinfo {author} {\bibfnamefont {M.}~\bibnamefont
  {Piarulli}}, \bibinfo {author} {\bibfnamefont {S.}~\bibnamefont {Pastore}},
  \bibinfo {author} {\bibfnamefont {R.~B.}\ \bibnamefont {Wiringa}}, \bibinfo
  {author} {\bibfnamefont {S.}~\bibnamefont {Brusilow}}, \ and\ \bibinfo
  {author} {\bibfnamefont {R.}~\bibnamefont {Lim}},\ }\href {\doibase
  10.1103/PhysRevC.107.014314} {\bibfield  {journal} {\bibinfo  {journal}
  {Phys. Rev. C}\ }\textbf {\bibinfo {volume} {107}},\ \bibinfo {pages}
  {014314} (\bibinfo {year} {2023})}\BibitemShut {NoStop}%
\end{thebibliography}%
\end{document}